\documentclass[12pt]{article}
\usepackage{amssymb,epsfig}

\def\baa             {\begin{array}}
\def\eaa             {\end{array}}
\def\beqaa           {\begin{eqnarray}}
\def\eeqaa           {\end{eqnarray}}
\def\beqaad          {\begin{eqnarray*}}
\def\eeqaad          {\end{eqnarray*}}

\def\bce             {\begin{center}}
\def\ece             {\end{center}}
\def\btabu           {\begin{tabular}}
\def\etabu           {\end{tabular}}

\def\noi             {\noindent}

\def\b               {\beta}

\def\t               {\theta}
\def\x               {\chi}

\def\ti              {\tilde}

\def\delr            {\!\stackrel{\leftrightarrow}{\partial^\mu}\!}

\def\sq              {\ti q}
\def\sqi             {\ti q_i^{}}

\def\st              {\ti t}
\def\sb              {\ti b}

\def\ch              {\ti \x^+}

\def\chp             {\ti \x^+}
\def\chpc            {\ti \x^{+c}}

\def\nt              {\ti \x^0}

\def\tW              {\t_W}

\def\tsq             {\t_{\ti q}}

\def\aik             {a_{ik}}
\def\bik             {b_{ik}}
\def\lij             {\ell_{ij}}
\def\kij             {k_{ij}}

\def\PL              {P_L^{}}
\def\PR              {P_R^{}}

\def\lag             {{\cal L}}

\def\R               {{\cal R}}
\def\Rsq             {{\cal R}^{\sq}}
\def\Rst             {{\cal R}^{\st}}
\def\Rsb             {{\cal R}^{\sb}}

\def\rzw             {\sqrt{2}}

\def\onehf            {{\textstyle \frac{1}{2} }}

\newcommand{\gsim}{\;\raisebox{-0.9ex}
           {$\textstyle\stackrel{\textstyle >}{\sim}$}\;}
\newcommand{\lsim}{\;\raisebox{-0.9ex}{$\textstyle\stackrel{\textstyle<}
           {\sim}$}\;}

\begin{document}

\hspace*{\fill}TGU-32

\hspace*{\fill}UWThPh-2003-26

\hspace*{\fill}ZU-TH 14/03

\hspace*{\fill}hep-ph/0311338

\vspace{1cm}
\begin{center}
\textbf{\Large Top Squarks and Bottom Squarks in the MSSM with Complex Parameters}

\vspace{10mm}
{\large
A.~Bartl$^a$, S.~Hesselbach$^a$, K.~Hidaka$^b$, T.~Kernreiter$^a$,
W.~Porod$^c$
}

\vspace{5mm}
\itshape
$^a$Institut f\"ur Theoretische Physik, Universit\"at Wien, A-1090
  Vienna, Austria\\
$^b$Department of Physics, Tokyo Gakugei University, Koganei, Tokyo
  184-8501, Japan\\
$^c$Institut f\"ur Theoretische Physik, Universit\"at Z\"urich,
  CH-8057 Z\"urich, Switzerland

\end{center}

\vspace{5mm}
\begin{abstract}

We present a phenomenological study of top squarks $(\ti t_{1,2})$ and
bottom squarks $(\ti b_{1,2})$ in the Minimal Supersymmetric
Standard Model (MSSM) with complex parameters $A_t, A_b, \mu$ and $M_1$.
In particular we focus on the CP phase dependence
of the branching ratios of $\tilde t_{1,2}$ and $\tilde b_{1,2}$ decays.
We give the formulae of the two-body decay widths and present
numerical results.
We find that the effect of the phases on the $\tilde t_{1,2}$ and
$\tilde b_{1,2}$ decays can be quite significant in a large region
of the MSSM parameter space. This could have important implications
for $\tilde t_{1,2}$ and $\tilde b_{1,2}$ searches and the MSSM
parameter determination in future collider experiments.
We have also estimated the accuracy expected in the
determination of the parameters of $\tilde{t}_i$ and $\tilde{b}_i$ by
a global fit of the measured masses, decay branching ratios and production
cross sections at $e^+ e^-$ linear colliders with polarized beams. 
Analysing two scenarios, we
find that the fundamental parameters apart from $A_t$
and $A_b$ can be determined with errors of 1\,\% to 2\,\%, assuming an
integrated luminosity of 1~ab$^{-1}$ and a sufficiently large c.m.s.\
energy to produce also the heavier $\tilde{t}_2$ and $\tilde{b}_2$
states. The parameter $A_t$ can be
determined  with an error of 2 -- 3\,\%, whereas the error on $A_b$ 
is likely to be of the order of 50\,\%.

\end{abstract}

\section{Introduction}

Supersymmetry (SUSY) is one of the most attractive and best
studied extensions of the standard model (SM) \cite{susy}. 
With SUSY the
hierarchy problem can be solved and the mass of the Higgs boson
can be stabilized against radiative corrections. While
this is certainly the main motivation, 
SUSY gives us the additional benefit of introducing potential new
sources of CP violation \cite{Dugan:1984qf,Masiero:xj}.
As the tiny amount of CP violation
in the SM is not sufficient to explain the baryon asymmetry
of the universe \cite{baryonasym}, the systematic study of all implications of the
complex SUSY parameters becomes absolutely necessary.

In the present paper we study the effects
of complex SUSY parameters on the phenomenology of
the scalar top quark and scalar bottom quark system. 
Analysing the properties of 3rd generation sfermions
is particularly interesting, because of the effects of
the large Yukawa couplings.
Their lighter mass eigenstates may be among the
light SUSY particles and they could be investigated  
at the Tevatron and at
$e^+ e^-$ linear colliders
\cite{baerTevatron}--\cite{Boos:2003vf}.
At LHC these states can be produced directly or in cascade decays of 
heavier SUSY particles \cite{Bartl:bu}--\cite{Paige:2003sh}.
Analyses of the decays of the
3rd generation sfermions $\tilde{t}_{1,2}$, $\tilde{b}_{1,2}$,
$\tilde{\tau}_{1,2}$ and $\tilde{\nu}_\tau$ in the 
Minimal Supersymmetric Standard Model (MSSM) with real
parameters have been performed in 
Refs.~\cite{stop2-stau2}--\cite{eberlRealSf}. 
Phenomenological studies of production and decays of the 3rd
generation sfermions at future $e^+ e^-$ linear colliders, again in
the real MSSM, have been made in 
Refs.~\cite{sfLCnojiri}--\cite{sferm}.

In the MSSM several SUSY breaking 
parameters and the higgsino mass parameter $\mu$ can be complex.
In a complete
phenomenological analysis of production and decays of third
generation sfermions one has to take into account that the SUSY
parameters $A_f$, $\mu$ and $M_i$  ($i=1,2,3$) are
complex in general, where $A_f$ is the trilinear scalar coupling parameter
of the sfermion $\tilde{f}_i$, and the $M_1$, $M_2$ and $M_3$ are 
the U(1), SU(2) and SU(3) gaugino mass parameters, respectively.
This means that one has to study the effects of the phases of the
parameters on all observables.

An unambiguous signal for the CP phases
would be provided by a measurement of a CP-odd observable.
For example, in the case of sfermion decays a rate 
asymmetry \cite{Aoki:1998cq} and
triple product correlations \cite{Bartl:2002hi,Bartl:2003ck} 
have been proposed as such observables.
However, since it may be difficult
to measure these CP-odd observables of the sfermions,
CP-even observables like decay branching ratios may also be 
suitable to obtain informations about the SUSY CP phases.
For example, the decay
branching ratios of the Higgs bosons depend strongly on the complex
phases of the $\tilde{t}$ and $\tilde{b}$ sectors 
\cite{Demir:1999hj}--\cite{ref4},
while those of the staus $\tilde{\tau}_{1,2}$ and $\tau$-sneutrino
$\tilde{\nu}_\tau$ can be quite sensitive to the phases of the stau
and gaugino-higgsino sectors \cite{CPslepton}.
Also the Yukawa couplings of the third generation sfermions are
sensitive to the SUSY phases at one-loop level \cite{Ibrahim:2003ca}.
Furthermore, explicit CP violation in the Higgs sector can be induced
by $\tilde{t}$ and $\tilde{b}$ loops
if the parameters $A_t$, $A_b$ and $\mu$ are complex
\cite{Demir:1999hj,ref3,ref2,feynhiggs}. It is found 
\cite{Demir:1999hj,ref4,ref3,Carena:2002bb} that these CP phase
effects could significantly influence the phenomenology of the Higgs
boson sector. 

The experimental upper bounds on the electric dipole moments (EDM's) of
electron, neutron and the $^{199}$Hg and $^{205}$Tl atoms may
impose constraints on the size of the SUSY CP phases 
\cite{smallphases,Barger:2001nu}. 
However, these constraints are highly model dependent. 
This means that the various SUSY CP phases need not necessarily be small.
For instance, if we adopt the MSSM 
and assume a cancellation mechanism \cite{edm},
it turns out that the phase of $\mu$
is restricted as $|\varphi_{\mu}|\lsim \pi/10$ while the phases
$\varphi_{A_f}$ of the $A_f$ parameters are not constrained.
On the other hand, 
the size of $|\varphi_{\mu}|$ is not constrained by the EDM's
in a model where the
masses of the first and second generation sfermions are large
(above the TeV scale) while the masses of the third
generation sfermions are small (below $1$~TeV) \cite{cohen}.
The restrictions on $\varphi_{\mu}$ due to the electron EDM can 
also be circumvented if lepton flavour violating terms are present
in the slepton sector \cite{Bartl:2003ju}.
Less restrictive constraints on the phases appear at 
two-loop level where 3rd generation sfermion loops can contribute
to the EDM's \cite{Pilaftsis}.

In this article we focus on the influence of the 
CP violating SUSY phases on the fermionic and bosonic
two-body decay branching ratios of 3rd generation squarks 
$\tilde t_{1,2}$ and $\tilde b_{1,2}$.
We use the MSSM as a general framework and we assume that the
parameters $A_t$, $A_b$, $\mu$ and $M_1$ are complex 
with phases $\varphi_{A_t}$, $\varphi_{A_b}$, $\varphi_\mu$ and
$\varphi_\mathrm{U(1)}$, respectively (taking $M_{2,3}$ real).
We neglect flavor changing CP
phases and assume that the squark mass matrices and trilinear
scalar coupling parameters are flavor diagonal.
We take into account the explicit CP violation in the Higgs sector.
If the top squark and bottom squark decay branching ratios show an appreciable
phase dependence, this would also affect the analyses of the various
gluino cascade decays such as those in \cite{nojirietal}.
In \cite{short} we have published first results of our study. In the
present paper we give the analytic expressions for the various decay
widths for the complex parameters and study in detail the phase dependences
of the branching ratios. 
We take into account the restrictions on the MSSM parameters
from the experimental data on the rare decay $b\to s\gamma$ \cite{bsgamma}. 
Furthermore, we give a theoretical estimate of the
precision expected for the determination of the complex top squark and
bottom squark parameters by measuring suitable observables including the
decay branching ratios in typical future collider experiments.

In Section 2 we give the formulae necessary to
calculate the $\ti t_i$ and $\ti b_i$ two-body decay widths
in the presence of CP phases.
In Section 3 we present our numerical results. 
In Section 4 we give a theoretical estimate how precisely the complex
top squark and bottom squark parameters can be determined at future
collider experiments.
We present our conclusions in Section 5.

\section{Squark masses, mixing and decay widths}

\subsection{Masses and mixing in squark sector}

The left-right mixing of the top squarks and bottom squarks is described by a
hermitian $2 \times 2$ mass matrix which in the basis 
$(\tilde{q}_L,\tilde{q}_R)$ reads
\begin{equation}
\mathcal{L}^{\tilde{q}}_M = - (\tilde{q}_L^{\dagger},\tilde{q}_R^{\dagger}) 
 \left(\begin{array}{cc} 
    M_{\tilde{q}_{LL}}^2 & M_{\tilde{q}_{LR}}^2\\
    M_{\tilde{q}_{RL}}^2 & M_{\tilde{q}_{RR}}^2
 \end{array}\right)
 \left(\begin{array}{c} \tilde{q}_L \\ \tilde{q}_R \end{array}\right),
\end{equation}
with
\begin{eqnarray}
M_{\tilde{q}_{LL}}^2 & = & M_{\tilde{Q}}^2 
  + (I^q_{3L} - e_q \sin^2 \theta_W) \cos 2\beta\; m_Z^2 + m_q^2, \\
M_{\tilde{q}_{RR}}^2 & = & M_{\tilde{Q'}}^2 
  + e_q \sin^2 \theta_W \cos 2\beta\; m_Z^2 + m_q^2, \\
M_{\tilde{q}_{RL}}^2 & = & (M_{\tilde{q}_{LR}}^2)^* = 
  m_q \left(A_q - \mu^* (\tan\beta)^{-2 I^q_{3L}}\right), \label{mLRterm}
\end{eqnarray}
where $m_q$, $e_q$ and $I^q_{3L}$ are the mass, electric charge and weak
isospin of the quark $q=b,t$. $\theta_W$ denotes the weak mixing angle,
$\tan\beta = v_2/v_1$ with $v_1$ ($v_2$) being the vacuum expectation
value of the Higgs field $H^0_1$ ($H^0_2$) and 
$M_{\tilde{Q'}}= M_{\tilde{D}}$ ($M_{\tilde{U}}$) for $q = b$ ($t$). 
$M_{\tilde{Q}}$, $M_{\tilde{D}}$, $M_{\tilde{U}}$, $A_b$ and $A_t$
are the soft SUSY-breaking parameters of the top squark and bottom squark system.
In the case of complex parameters $\mu$ and $A_q$ the off-diagonal elements
$M_{\tilde{q}_{RL}}^2 = (M_{\tilde{q}_{LR}}^2)^*$ are also complex
with the phase
\begin{equation} \label{defphisquark}
\varphi_{\tilde{q}} = \arg\left[M_{\tilde{q}_{RL}}^2\right] 
  = \arg\left[A_q - \mu^* (\tan\beta)^{-2 I^q_{3L}}\right].
\end{equation}
The mass eigenstates are
\begin{equation}
{\tilde{q}_1 \choose \tilde{q}_2} = 
  \mathcal{R}^{\tilde{q}} {\tilde{q}_L \choose \tilde{q}_R}
\end{equation}
with the $\ti q$-mixing matrix
\begin{equation}
\mathcal{R}^{\tilde{q}} =
\left(\begin{array}{cc}
  e^{i\varphi_{\tilde{q}}} \cos\theta_{\tilde{q}} & \sin\theta_{\tilde{q}}\\
  -\sin\theta_{\tilde{q}} & e^{-i\varphi_{\tilde{q}}} \cos\theta_{\tilde{q}}
\end{array}\right),
\end{equation}
\begin{equation} \label{eq:thetasquark}
\cos\theta_{\tilde{q}} =
 \frac{-|M_{\tilde{q}_{LR}}^2|}{\sqrt{|M_{\tilde{q}_{LR}}^2|^2
  + (m_{\tilde{q}_1}^2 - M_{\tilde{q}_{LL}}^2)^2}},\quad
\sin\theta_{\tilde{q}} =
 \frac{M_{\tilde{q}_{LL}}^2 - m_{\tilde{q}_1}^2}{\sqrt{|M_{\tilde{q}_{LR}}^2|^2
  + (m_{\tilde{q}_1}^2 - M_{\tilde{q}_{LL}}^2)^2}}
\end{equation}
and the mass eigenvalues
\begin{equation} \label{eq:msquark}
m_{\tilde{q}_{1,2}}^2 = \frac{1}{2} \left(
  M_{\tilde{q}_{LL}}^2 + M_{\tilde{q}_{RR}}^2 \mp
  \sqrt{(M_{\tilde{q}_{LL}}^2 - M_{\tilde{q}_{RR}}^2)^2 +
    4 |M_{\tilde{q}_{LR}}^2|^2}\right),\hspace{1cm}
   m_{\tilde{q}_1} < m_{\tilde{q}_2}\:.
\end{equation}

\subsection{Fermionic decay widths of $\ti t_i$ and $\ti b_i$}

\noi 
In the following we give the formulae necessary to calculate
the two-body decay widths of $\ti t_i$ and $\ti b_i$
into charginos and neutralinos in the presence of the CP phases.
The $b-\ti t_i-\ti\chi^{\pm}_k$ and $t-\ti b_i-\ti\chi^{\pm}_k$ 
couplings are defined by
\begin{equation}
  \lag_{q\sq\ch} 
  = g\,\bar t\,
    (\ell_{ij}^{\ti b}\,\PR + k_{ij}^{\ti b}\,\PL)\,\chp_j\,\sb_i^{} + 
    g\,\bar b\,
    (\ell_{ij}^{\ti t}\,\PR + k_{ij}^{\ti t}\,\PL)\,\chpc_j\,\st_i^{}+
    {\rm h.c.}
\label{eq:qsqch}
\end{equation}
with
\begin{equation}
  \hspace*{1mm}
  P_L = \frac{1}{2}(1 - \gamma_5), \qquad P_R = \frac{1}{2}(1 + \gamma_5),
\end{equation}
\begin{equation}\label{eq:ltij}
  \hspace*{21mm}
  \lij^{\st} = -{\Rst}^{\ast}_{i1} V_{j1}+
Y_t {\Rst}^{\ast}_{i2} V_{j2}, \qquad
  \kij^{\st} = {\Rst}^{\ast}_{i1}\,Y_b\,U_{j2}^{\ast},
\end{equation}
\begin{equation}
  \hspace*{21mm}
  \lij^{\sb} = -{\Rsb}^{\ast}_{i1} U_{j1}+
Y_b {\Rsb}^{\ast}_{i2} U_{j2}, \qquad
  \kij^{\sb} = {\Rsb}^{\ast}_{i1}\,Y_t\,V_{j2}^{\ast}
\end{equation}
and
\begin{equation}
  \hspace*{21mm}
  Y_t = \frac{m_t}{\sqrt{2}\:m_W\sin\b}, \hspace{8mm} 
  Y_b = \frac{m_b}{\sqrt{2}\:m_W\cos\b},
\end{equation}
where $g$ is the $\mathrm{SU(2)}_L$ gauge coupling and
the 2$\times$2 chargino mixing matrices $U$ and $V$ are defined in 
Eq. (\ref{dia}).

\noi 
The $q-\ti q_i-\ti\chi^0_k$ couplings ($q=t,b$) are defined by
\begin{eqnarray}
  \lag_{q\sq\nt} 
  &=& g\,\bar{q}\,(a_{ik}^{\sq}\,\PR + b_{ik}^{\sq}\,\PL)\,\nt_k\,\sqi
    + {\rm h.c.}\ ,
\end{eqnarray}        
with
\begin{equation} \label{eq:abneut}
  \hspace*{20mm}
  \aik^{\sq} = \sum^2_{n=1}\, (\Rsq_{in})^{\ast}\,{\cal A}_{kn}^q, \qquad
  \bik^{\sq} = \sum^2_{n=1}\, (\Rsq_{in})^{\ast}\,{\cal B}_{kn}^{\,q}\ ,
\end{equation}
where
\begin{equation}\label{eq:abneut1}
  \hspace*{20mm}
  {\cal A}_k^q = {f_{Lk}^q \choose h_{Rk}^q}, \qquad
  {\cal B}_k^q = {h_{Lk}^q \choose f_{Rk}^q},
\end{equation}
\begin{eqnarray}
  \hspace*{2cm}
  f_{Lk}^t & = &
- {1\over \rzw}\Bigl( N_{k2}+{1\over 3}\tan\tW N_{k1}\Bigr)\ ,
    \hspace{-1cm} \label{eq:fLkt}\nonumber \\
  f_{Rk}^t & = &
    {2\rzw\over3}\tan\tW N^{\ast}_{k1}\ , \nonumber \\
  h_{Lk}^t & = & (h_{Rk}^t)^{\ast} = - Y_t N^{\ast}_{k4}\ ,
    \label{eq:htLk}
\end{eqnarray}
and
\begin{eqnarray}
  \hspace*{2cm}
  f_{Lk}^b & = &
 {1\over \rzw}\Bigl( N_{k2}-{1\over 3}\tan\tW N_{k1}\Bigr)\ ,
    \hspace{-1cm}\label{eq:fLkb}\nonumber \\
  f_{Rk}^b & = &
    -{\rzw\over3}\tan\tW N^{\ast}_{k1}\ , \nonumber \\
  h_{Lk}^b & = & (h_{Rk}^b)^{\ast} = -Y_b N^{\ast}_{k3}\ .
    \label{eq:hbLk}
\end{eqnarray}
The 4$\times$4 neutralino mixing matrix $N$ is defined in
Eq.~(\ref{eq:mixN}).  

\noi
The partial decay widths of $\ti q_i$ ($\ti q_i=\ti t_i,\ti b_i$) 
into fermionic final states then read
\begin{eqnarray}
\Gamma(\tilde{q}_i \to q' + \tilde{\chi}^\pm_k) & = &
 \frac{g^2 
  \lambda^\frac{1}{2}(m_{\tilde{q}_i}^2,m_{q'}^2,m_{\tilde{\chi}^\pm_k}^2)}
  {16 \pi m_{\tilde{q}_i}^3} \times \nonumber \\
 & & \left[\left(|k_{ik}^{\tilde{q}}|^2 + |\ell_{ik}^{\tilde{q}}|^2 \right)
         \left(m_{\tilde{q}_i}^2 - m_{q'}^2 - m_{\tilde{\chi}^\pm_k}^2 \right)
          - 4 \mathrm{Re}(k_{ik}^{\tilde{q}*} \ell_{ik}^{\tilde{q}})
                m_{q'} m_{\tilde{\chi}^\pm_k} \right]
 \label{eq:gamtC}
\end{eqnarray}
and
\begin{eqnarray}
\Gamma(\tilde{q}_i \to q + \tilde{\chi}^0_k) & = &
 \frac{g^2 
  \lambda^\frac{1}{2}(m_{\tilde{q}_i}^2,m_q^2,m_{\tilde{\chi}^0_k}^2)}
  {16 \pi m_{\tilde{q}_i}^3} \times \nonumber \\
 & & \left[\left(|a_{ik}^{\tilde{q}}|^2 + |b_{ik}^{\tilde{q}}|^2 \right)
         \left(m_{\tilde{q}_i}^2 - m_q^2 - m_{\tilde{\chi}^0_k}^2 \right)
          - 4 \mathrm{Re}(a_{ik}^{\tilde{q}*} b_{ik}^{\tilde{q}})
                m_q m_{\tilde{\chi}^0_k} \right]
 \label{eq:gamtN}
\end{eqnarray}
with $\lambda(x,y,z) = x^2 + y^2 + z^2 - 2(xy + xz + yz)$.

\subsection{Bosonic decay widths of $\ti t_i$ and $\ti b_i$}

\noi
Here we show the couplings relevant for the two-body
decays of $\ti t_i$ and $\ti b_i$ into gauge and Higgs bosons.
The $\ti q_i-\ti q'_j-W^{\pm}$ couplings are defined by
\begin{equation} 
  \lag_{\ti{q}\ti{q}' W}
  = -ig\, 
  (A^W_{\sb_i\st_j}\,W^+_\mu\,\st_j^{\dagger}\delr\sb_i^{} + 
   A^W_{\st_i\sb_j}\,W^-_\mu\,\sb_j^{\dagger}\delr\st_i^{}) 
\label{eq:csqW}
\end{equation}
with
\begin{equation}
  A^W_{\sb_i\st_j} = (A^W_{\st_j\sb_i})^\ast =
    \frac{1}{\sqrt{2}} {\R_{i1}^{\sb}}^{\!\!\ast} \R_{j1}^{\st} \,.
\end{equation}
 
\noi
The $\ti q_i-\ti q_j-Z$ interaction Lagrangian reads
\begin{equation}
  \lag_{\ti{q}\ti{q} Z} 
  = -ig\: B^Z_{ij} \, Z_\mu\,\sq_j^{\dagger}\delr\sqi \,
\label{eq:L-ssZ}
\end{equation}
with
\begin{equation}
  B^Z_{ij} = \frac{1}{\cos\tW} \left( \begin{array}{cc}
    I^{q}_{3L}\,\cos^2\tsq - e_q \sin^2\tW 
         & -\onehf\, I^{q}_{3L}\,\sin 2\tsq \,e^{-i\varphi_{\sq}}\\[1mm]
    -\onehf\, I^{q}_{3L}\,\sin 2\tsq \,e^{i\varphi_{\sq}}
         & I^{q}_{3L}\,\sin^2\tsq - e_q \sin^2\tW
  \end{array} \right) .
\label{eq:cij}
\end{equation}

\noi 
The $\ti q_i-\ti q'_j-H^{\pm}$ couplings are defined by
\begin{equation}
\lag_{\ti{q}\ti{q} H^\pm} = g \left(
  C^H_{\st_j\sb_i} H^+ \st_j^{\dagger}\,\sb_i + 
  C^H_{\sb_j\st_i} H^- \sb_j^{\dagger}\,\st_i \right)
\end{equation}
with
\begin{equation} \label{eq:G}
C^H_{\st_i\,\sb_j} = (C^H_{\sb_j\,\st_i})^\ast  = \frac{1}{\rzw\,m_W^{}}
  (\R^{\st} G {\R^{\sb}}^\dagger)_{ij}
\end{equation}
and
\begin{equation}
 G = \left(\! \begin{array}{cc}
   m_b^2\tan\b + m_t^2\cot\b - m_W^2\sin 2\b 
     & m_b\,(|A_b| e^{-i\varphi_{A_b}} \tan\b + |\mu| e^{i\varphi_\mu}) \\[3mm]
   m_t\,(|A_t| e^{i\varphi_{A_t}} \cot\b + |\mu| e^{-i\varphi_\mu}) 
     & \displaystyle\frac{2 m_t m_b}{\sin 2\b}
\end{array}\! \right) .
\label{eq:G1}
\end{equation}

\noi 
For the couplings of squarks to neutral Higgs bosons we have the 
Lagrangian

\begin{equation}
  \lag_{\ti{q}\ti{q}H} = - g \, C(\sq_k^{\dagger} H_i \sq_j) \,
     \sq_k^{\dagger} H_i \sq_j \quad (k,j=1,2)
\end{equation}
with 
\begin{equation}
C(\sq_k^{\dagger} H_i \sq_j) = \Rsq\cdot
\left(\begin{array}{ccc}
C(\sq_{L}^{\dagger} H_i \sq_{L}) & C(\sq_{L}^{\dagger} H_i \sq_{R})\\[5mm]
 C({\sq_{R}}^{\dagger} H_i \sq_{L}) & C({\sq_{R}}^{\dagger} H_i \sq_{R})
\end{array}\right)\cdot {\Rsq}^{\dag} ,
\label{eq:Hijcoup}
\end{equation}
where for $\ti q=\ti t$ 
\begin{equation}
C(\st_{L}^{\dagger} H_i \st_{L})=\frac{m_t^2}{m_W
\sin\beta} O_{2i} 
+\frac{m_Z}{\cos\theta_W}\left(\frac{1}{2}-\frac{2}{3}\sin^2\theta_W\right)
\left(\cos\beta O_{1i} 
-\sin\beta O_{2i}\right),
\label{eq:HLLcoupt}
\end{equation}

\begin{equation}
C(\st_{R}^{\dagger} H_i \st_{R})=\frac{m_t^2}{m_W
\sin\beta} O_{2i} 
+\frac{2 m_Z}{3\cos\theta_W} \sin^2\theta_W(\cos\beta O_{1i}
-\sin\beta O_{2i}),
\label{eq:HRRcoupt}
\end{equation}

\begin{eqnarray}
C(\st_{L}^{\dagger} H_i \st_{R})&=& \frac{m_t}{2 m_W
\sin\beta}\{-i \left(\cos\beta |A_t| e^{-i\varphi_{A_t}}
+\sin\beta |\mu|e^{i\varphi_{\mu}}\right) O_{3i}
\nonumber\\[3mm]
&&{}-\left(|\mu|e^{i\varphi_{\mu}} O_{1i}- 
|A_t| e^{-i\varphi_{A_t}} O_{2i}\right)\},
\label{eq:HLRcoupt}
\end{eqnarray}

\begin{equation}
C(\st_{R}^{\dagger} H_i \st_{L})=\lbrack C(\st_{L}^{\dagger} H_i \st_{R})
\rbrack^{\ast},
\label{eq:HRLcoupt}
\end{equation}

while for $\ti q=\ti b$
\begin{equation}
C(\sb_{L}^{\dagger} H_i \sb_{L})=\frac{m_b^2}{m_W
\cos\beta} O_{1i} 
-\frac{m_Z}{\cos\theta_W}\left(\frac{1}{2}-\frac{1}{3}\sin^2\theta_W\right)
\left(\cos\beta O_{1i} 
-\sin\beta O_{2i}\right),
\label{eq:HLLcoupb}
\end{equation}

\begin{equation}
C(\sb_{R}^{\dagger} H_i \sb_{R})=\frac{m_b^2}{m_W
\cos\beta} O_{1i} 
-\frac{m_Z}{3\cos\theta_W} \sin^2\theta_W(\cos\beta O_{1i}
-\sin\beta O_{2i}),
\label{eq:HRRcoupb}
\end{equation}

\begin{eqnarray}
C(\sb_{L}^{\dagger} H_i \sb_{R})&=& \frac{m_b}{2 m_W
\cos\beta}\{-i \left(\sin\beta |A_b| e^{-i\varphi_{A_b}}
+\cos\beta |\mu|e^{i\varphi_{\mu}}\right) O_{3i}
\nonumber\\[3mm]
&&{}-\left(|\mu|e^{i\varphi_{\mu}} O_{2i}- 
|A_b| e^{-i\varphi_{A_b}} O_{1i}\right)\},
\label{eq:HLRcoupb}
\end{eqnarray}

\begin{equation}
C(\sb_{R}^{\dagger} H_i \sb_{L})=\lbrack C(\sb_{L}^{\dagger} H_i \sb_{R})
\rbrack^{\ast}.
\end{equation}
Here the $3 \times 3$ neutral Higgs mixing matrix $O$ is defined in
Eq.~(\ref{eq:nhiggsmix}).

The partial decay widths for $\ti q_i=\ti t_i,\ti b_i$ 
into bosonic final states are then of the following forms:
\begin{equation}
\Gamma(\tilde{q}_i \to W^\pm + \tilde{q}'_j) =
  \frac{g^2 |A^W_{\tilde{q}_i \tilde{q}'_j}|^2
  \lambda^\frac{3}{2}(m_{\tilde{q}_i}^2,m_W^2,m_{\tilde{q}'_j}^2)}
    {16 \pi m_W^2 m_{\tilde{q}_i}^3} \;,
\end{equation}

\begin{equation}
\Gamma(\tilde{q}_2 \to Z + \tilde{q}_1) =
  \frac{g^2 |B^Z_{21}|^2
  \lambda^\frac{3}{2}(m_{\tilde{q}_2}^2,m_Z^2,m_{\tilde{q}_1}^2)}
    {16 \pi m_Z^2 m_{\tilde{q}_2}^3} \;,
\end{equation}

\begin{equation}
\Gamma(\tilde{q}_i \to H^\pm + \tilde{q}'_j) =
  \frac{g^2 |C^H_{\tilde{q}'_j \tilde{q}_i}|^2
  \lambda^\frac{1}{2}(m_{\tilde{q}_i}^2,m_{H^\pm}^2,m_{\tilde{q}'_j}^2)}
    {16 \pi m_{\tilde{q}_i}^3} \;,
\end{equation}

\begin{equation} \label{eq:gamneuthiggs}
\Gamma(\tilde{q}_2 \to H_i + \tilde{q}_1) =
  \frac{g^2 |C(\sq_1^{\dagger} H_i \sq_2)|^2
  \lambda^\frac{1}{2}(m_{\tilde{q}_2}^2,m_{H_i}^2,m_{\tilde{q}_1}^2)}
    {16 \pi m_{\tilde{q}_2}^3} \;.
\end{equation}

\section{Numerical results}

Before presenting numerical results, we briefly comment
on the CP phase dependence of the $\ti q_i\bar{\ti q_j}$ pair production
cross sections.
The reaction $e^+ e^- \to \tilde{q}_i \bar{\tilde{q}}_j$ 
($\tilde{q}_i = \tilde{t}_i, \tilde{b}_i$) proceeds via
$\gamma$ and $Z$ exchange in the $s$-channel. The 
$Z \tilde{q}_i \tilde{q}_j$ couplings are defined in
Eqs.~(\ref{eq:L-ssZ}) and (\ref{eq:cij}).
The tree-level cross sections \cite{sfLCbartl,sferm} of the reactions
$e^+ e^- \to \tilde{q}_i \bar{\tilde{q}}_j$ do not explicitly depend
on the phases $\varphi_\mu$ and $\varphi_{A_q}$.
In the case of the reaction 
$e^+ e^- \to \tilde{q}_i \bar{\tilde{q}}_i$, $i=1,2$, the couplings
$Z \tilde{q}_i \tilde{q}_i$ are real.
In $e^+ e^- \to \tilde{q}_1 \bar{\tilde{q}}_2$ only $Z$ exchange
contributes and consequently the phase $\varphi_{\sq}$ drops out in
the matrix element squared.
The tree-level cross sections depend only on the mass eigenvalues
$m_{\tilde{q}_{1,2}}$ and on the mixing angle
$\cos^2\theta_{\tilde{q}}$. Therefore, they depend only implicitly on the
phases via the $\cos(\varphi_\mu + \varphi_{A_q})$ dependence of
$m_{\tilde{q}_{1,2}}$ and $\theta_{\tilde{q}}$
(Eqs.~(\ref{eq:thetasquark}) and (\ref{eq:msquark})).
One-loop corrections to the cross sections have been
calculated in \cite{loopcrosssection} for real parameters. In the
energy range considered here they are of the order of 10\,\%.
For complex parameters they are expected to be of the
same order of magnitude. Therefore we further expect that the direct
influence of the phases on the cross sections as caused by one-loop
corrections would be within a few percent. These phase effects on the
cross sections would be much smaller than those on the tree-level decay
widths studied in this paper.

In the following we will present numerical results for the phase
dependences of the $\st_i$ and $\sb_i$ partial decay widths and branching
ratios. We calculate the partial decay widths in Born approximation
according to the expressions given in the preceding section. In some
cases the one-loop SUSY QCD corrections are important. The analyses of
\cite{stop1, mbrun, mbrun2} suggest that a significant part of the
one-loop SUSY QCD
corrections to certain partial widths of $\ti t_i$ and $\ti b_i$ decays
(where the bottom Yukawa coupling $g Y_b$ is involved)
can be
incorporated by using an appropriately corrected bottom quark mass. In
this spirit we calculate the tree-level widths of the $\st_i$ and
$\sb_i$ decays by using on-shell masses for the kinematic 
terms (such as a phase space factor) and by taking running
$t$ and $b$ quark masses for the Yukawa couplings $g Y_{t,b}$.
For definiteness we take
$m_t^\mathrm{run}(m_Z) = 150$~GeV,
$m_t^\textrm{\scriptsize on-shell} = 175$~GeV,
$m_b^\mathrm{run}(m_Z) = 3$~GeV and
$m_b^\textrm{\scriptsize on-shell} = 5$~GeV.
This approach leads to an ``improved'' Born approximation which takes
into account an essential part
of the one-loop SUSY QCD corrections to the $\st_i$ and $\sb_i$
partial decay widths and predicts their phase dependences more accurately
than the ``naive'' tree-level calculation.
The inclusion of the full one-loop corrections to the partial decay
widths of $\st_i$ and $\sb_i$ is beyond the scope of the present paper.
One-loop corrections to partial decay widths of $\st_i$ and
$\sb_i$ have been given in \cite{loopdecay, Guasch:2002ez} 
for real MSSM parameters and are of the order of 10\,\%. We expect
that for complex parameters they are of the same order of magnitude.
In the calculation of the CP violating effects in the neutral Higgs
sector we take the program FeynHiggs2.0.2 of \cite{feynhiggs},
which includes the full one-loop corrections to the mass
eigenvalues and mixing matrix of the neutral Higgs bosons for complex
parameters.
For comparison we have also used the program cph.f of \cite{ref3}. We have
found agreement between the results obtained with cph.f and the one-loop
version of FeynHiggs2.0.2. There are small numerical differences
between the results of cph.f and the two-loop version of FeynHiggs2.0.2.

In the numerical analysis we impose the following conditions in order
to fulfill the experimental and theoretical constraints:
\renewcommand{\labelenumi}{(\roman{enumi})}
\begin{enumerate}
\item $m_{\tilde{\chi}^\pm_1} > 103$~GeV, 
  $m_{\tilde{\chi}^0_1} > 50$~GeV,
  $m_{\st_1,\sb_1} > 100$~GeV, $m_{\st_1,\sb_1} > m_{\tilde{\chi}^0_1}$,

\item for incorporating the experimental bound on the mass of
 the lightest Higgs boson $H_1$ we use Fig.~4 of \cite{Heister:2001kr},
 replacing $m_h$ by $m_{H_1}$ and $\sin^2(\beta - \alpha)$ by 
 $(O_{11} \cos\beta + O_{21} \sin\beta)^2$,\footnote{Note that
  $O_{11}\sim -\sin\alpha$, $O_{21}\sim \cos\alpha$ and $H_1 \sim h$
  for $m_{H^\pm} \gg m_Z$.}

\item $2.0 \times 10^{-4} < B(b \to s \gamma) < 4.5 \times 10^{-4}$
 \cite{bsgamma} assuming the Kobayashi-Maskawa mixing also
 for the squark sector,

\item $\Delta\rho(\ti t-\ti b) < 0.0012$ \cite{deltarho},

\item $|A_t|^2 < 3 (M_{\tilde{Q}}^2 + M_{\tilde{U}}^2 + m_2^2)$,
 $|A_b|^2 < 3 (M_{\tilde{Q}}^2 + M_{\tilde{D}}^2 + m_1^2)$ with\\
 $m_1^2 = (m_{H^\pm}^2 + m_Z^2 \sin^2\theta_W)\sin^2\beta -
   \frac{1}{2} m_Z^2$,
 $m_2^2 = (m_{H^\pm}^2 + m_Z^2 \sin^2\theta_W)\cos^2\beta -
   \frac{1}{2} m_Z^2$.

\end{enumerate}
Conditions (i) and (ii) are imposed to satisfy the experimental mass
bounds from LEP \cite{Heister:2001kr,LEPsusy,LEPhiggs}. Note that the CP
violation effect reduces the $Z-Z-H_1$ coupling because $H_1$ can have an
admixture of the CP-odd Higgs state $a$. 
The vertical axis of Fig.~4 of \cite{Heister:2001kr} describes the
$Z-Z-h$ coupling in the case of the MSSM with real parameters, which
is reduced by a factor $\sin^2(\beta - \alpha)$ in comparison to the SM.
The CP violating
effects can easily be included by using
$(O_{11} \cos\beta + O_{21} \sin\beta)^2$ instead of 
$\sin^2(\beta - \alpha)$.
For the calculation of the $b\to s\gamma$ width in
condition (iii) we use the formula
of \cite{Bertolini:1990if} including the O($\alpha_s$) corrections as given 
in \cite{Kagan:1998ym}.
(iv) constrains $\mu$ and $\tan\b$ (in the squark sector). 
(v) is the approximate necessary condition for the tree-level 
vacuum stability \cite{Derendinger-Savoi}.

Inspired by the gaugino mass unification we take
$|M_1|=5/3\tan^2\theta_W M_2$ and 
$m_{\ti g}=(\alpha_s(m_{\ti g})/\alpha_2) M_2$
with $m_{\ti g}=M_3$. In the numerical study for $\ti t_{1,2}$
decays we take 
$\tan\beta, M_2,$ $m_{\ti t_1}, m_{\ti t_2}, m_{\ti b_1},
|A_t|, |A_b|, |\mu|, \varphi_{A_t}, \varphi_{A_b}, \varphi_{\mu},
\varphi_\mathrm{U(1)}$ and $m_{H^\pm}$ 
as input parameters, where $m_{\ti t_{1,2}}$
and $m_{\ti b_{1,2}}$ are the on-shell squark masses.
From these input parameters we first calculate $M_{\tilde{Q}}$
and $M_{\tilde{U}}$ according to the formulae
\begin{eqnarray}
M_{\tilde{Q}}^2 & = & \frac{1}{2} \left( 
  m_{\ti t_1}^2 + m_{\ti t_2}^2 \pm 
  \sqrt{(m_{\ti t_2}^2 - m_{\ti t_1}^2)^2 - 
    4 m_t^2 \left|A_t - \mu^* \cot\beta\right|^2}\right) \nonumber \\
 && {}- (\frac{1}{2} - \frac{2}{3} \sin^2 \theta_W) \cos 2\beta\; m_Z^2 
     - m_t^2\,,\\
M_{\tilde{U}}^2 & = & \frac{1}{2} \left( 
  m_{\ti t_1}^2 + m_{\ti t_2}^2 \mp
  \sqrt{(m_{\ti t_2}^2 - m_{\ti t_1}^2)^2 - 
    4 m_t^2 \left|A_t - \mu^* \cot\beta\right|^2}\right) \nonumber \\
 && {}- \frac{2}{3} \sin^2 \theta_W \cos 2\beta\; m_Z^2
     - m_t^2\,.
\end{eqnarray}
We resolve the sign ambiguity by assuming either
$M_{\tilde{Q}}\geq M_{\tilde{U}}$ or
$M_{\tilde{Q}}<M_{\tilde{U}}$:
upper (lower) signs correspond to the case
$M_{\tilde{Q}}\geq M_{\tilde{U}}$ ($M_{\tilde{Q}}<M_{\tilde{U}}$).%
\footnote{The hierarchy is determined
by the $\st_i$ mixing angle $\theta_{\st}$, which can be determined by
cross section measurements with polarized beams
\cite{sfLCbartl,sferm}.}
With Eq.~(\ref{eq:thetasquark}) this uniquely fixes the mixing angle
$\theta_{\st}$. Next we calculate $M_{\tilde{D}}$ using
$M_{\tilde{Q}}$ and $m_{\tilde{b}_1}$ and then $m_{\tilde{b}_2}$  and the
mixing angle $\theta_{\sb}$ as well as the mass eigenvalues and mixing
matrices of the charginos, neutralinos and the neutral Higgs bosons.
For $\ti b_{1,2}$ decays we take the same input parameters 
with $m_{\ti t_2}$ replaced by $m_{\ti b_2}$
and proceed in the analogous way by interchanging 
$M_{\tilde{U}} \leftrightarrow M_{\tilde{D}}$.

\subsection{Top squark decays}

In this subsection we present numerical results for the dependence of
the  $\tilde{t}_1$ and $\tilde{t}_2$ partial decay widths on $\varphi_{A_t}$,
$\varphi_{A_b}$ and $\varphi_\mu$.
In order not to vary too many parameters we fix 
$(m_{\tilde{t}_1}, m_{\tilde{t}_2}, m_{\tilde{b}_1})=
(350, 700, 170)$~GeV $\lbrack(350, 800, 170)~{\rm GeV} \rbrack$
in the plots for the $\tilde{t}_1 \lbrack\tilde{t}_2\rbrack$ decays.
We have selected the parameters in this subsection such that
fermionic as well as bosonic decays are allowed at the same time. In
particular, the choice $m_{\sb_1} = 170$~GeV has been made to allow
the decays $\st_1 \to \sb_1 W^+$ and $\st_1 \to \sb_1 H^+$.
We consider the cases $M_{\tilde{Q}}\geq M_{\tilde{U}}$
and $M_{\tilde{Q}}<M_{\tilde{U}}$, calculating the values of
$M_{\tilde{Q}}$, $M_{\tilde{U}}$ and $M_{\tilde{D}}$ corresponding to 
$m_{\tilde{t}_1}$, $m_{\tilde{t}_2}$ and $m_{\tilde{b}_1}$ for each case,
as explained above.

\begin{figure}[p]
\centering
\begin{picture}(16,16.9)

\put(-.1,11.6){\epsfig{file=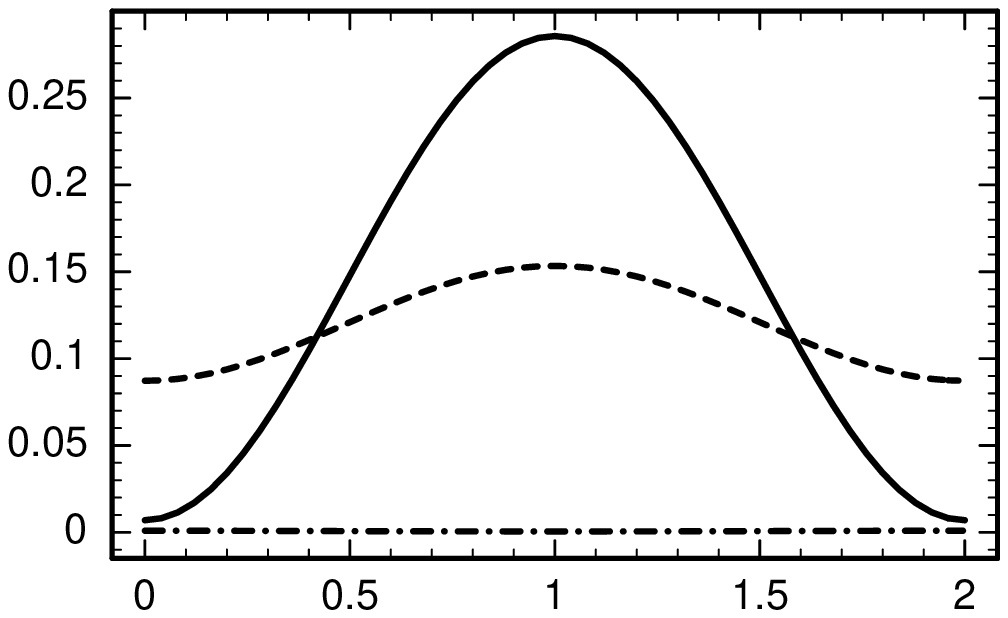,scale=.76}}
\put(0.1,16.5){$\Gamma/$GeV}
\put(6.5,11.3){$\varphi_{A_t}/\pi$}
\put(2,16.6){(a) $M_{\tilde{Q}}>M_{\tilde{U}}$, $|\mu|=350$~GeV}

\put(8.3,11.6){\epsfig{file=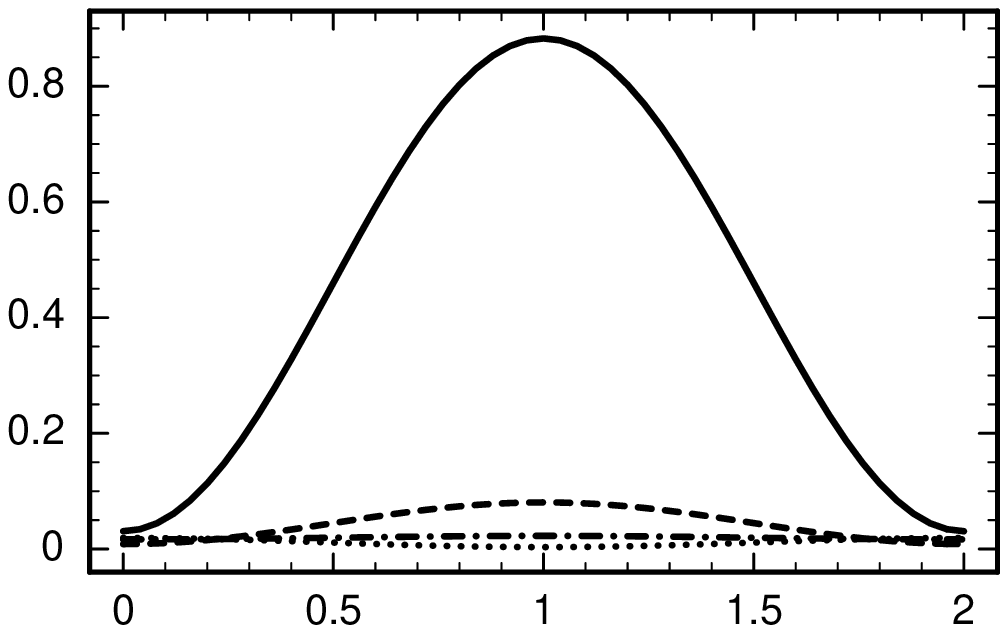,scale=.75}}
\put(8.4,16.5){$\Gamma/$GeV}
\put(14.8,11.3){$\varphi_{A_t}/\pi$}
\put(10.3,16.6){(d) $M_{\tilde{Q}}<M_{\tilde{U}}$, $|\mu|=250$~GeV}

\put(0,6){\epsfig{file=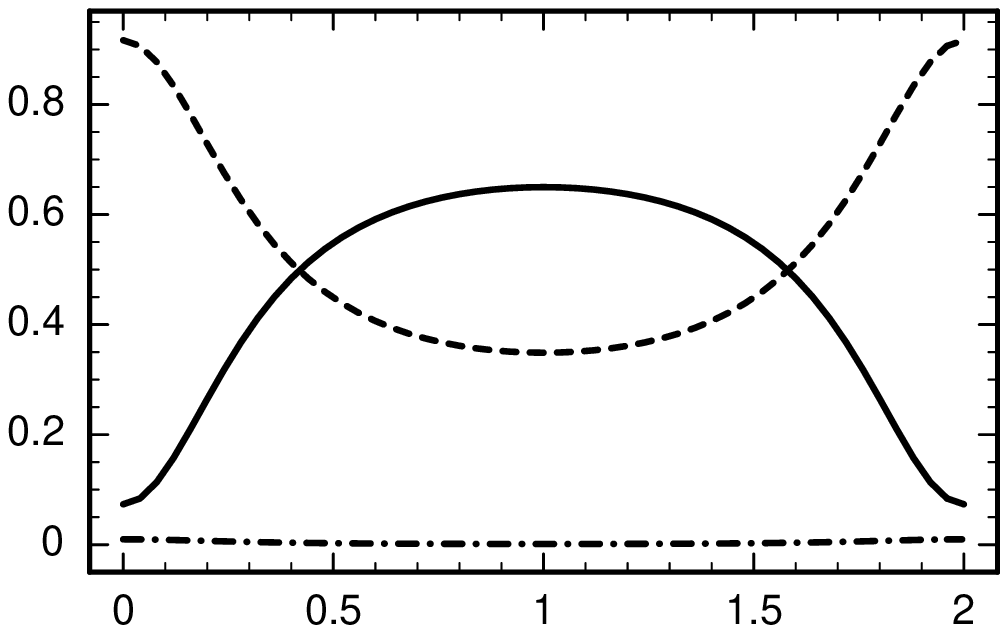,scale=.75}}
\put(0.1,10.6){$B$}
\put(6.5,5.7){$\varphi_{A_t}/\pi$}
\put(0.7,11.1){(b) $M_{\tilde{Q}}>M_{\tilde{U}}$, $|\mu|=350$~GeV}

\put(8.3,6){\epsfig{file=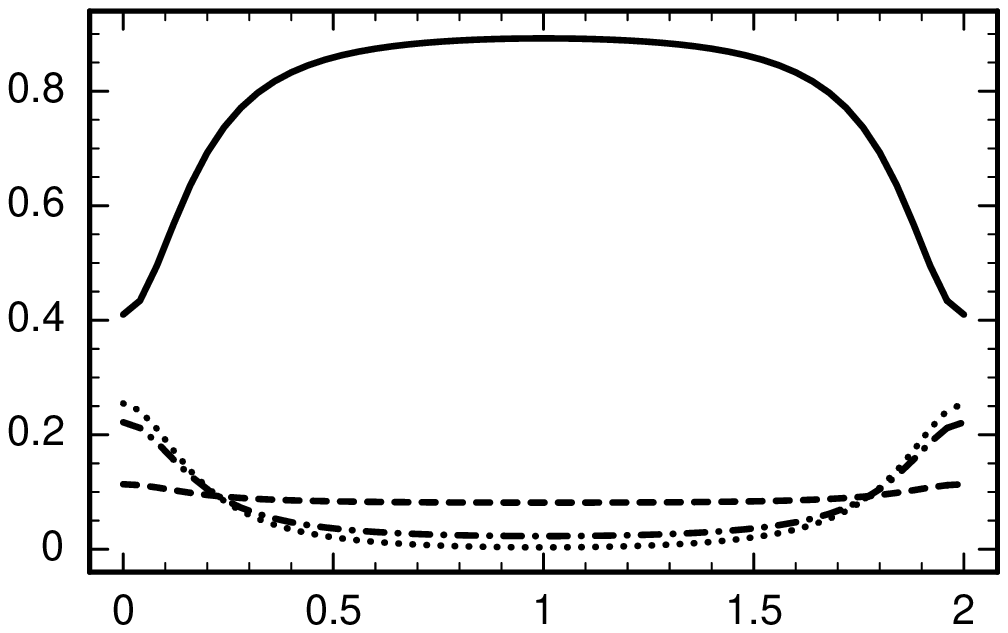,scale=.75}}
\put(8.4,10.6){$B$}
\put(14.8,5.7){$\varphi_{A_t}/\pi$}
\put(9,11.1){(e) $M_{\tilde{Q}}<M_{\tilde{U}}$, $|\mu|=250$~GeV}

\put(0,0.4){\epsfig{file=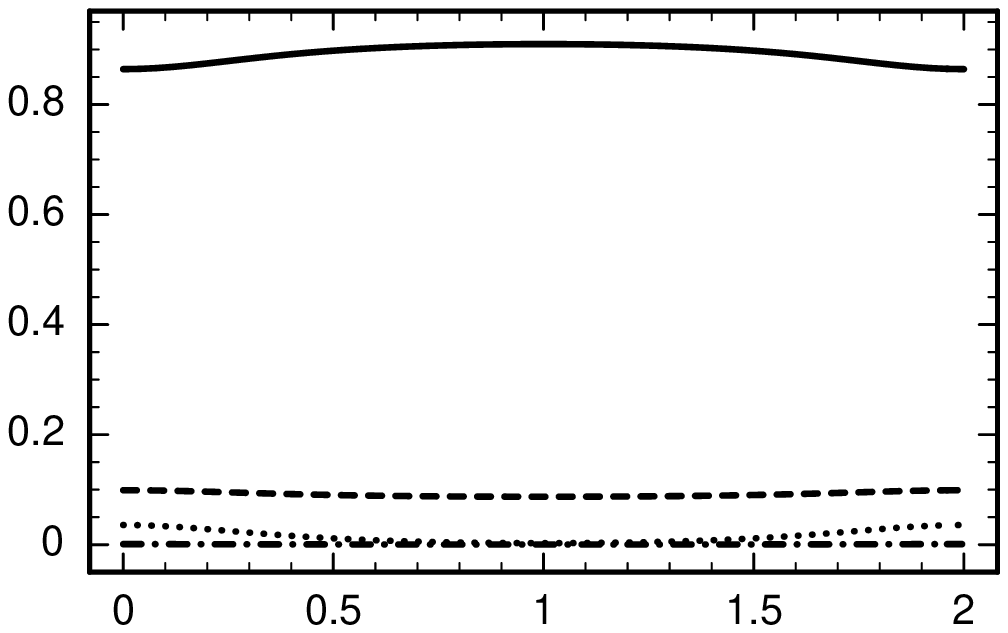,scale=.75}}
\put(0.1,5){$B$}
\put(6.5,0.1){$\varphi_{A_t}/\pi$}
\put(0.7,5.5){(c) $M_{\tilde{Q}}>M_{\tilde{U}}$, $|\mu|=250$~GeV}

\put(8.3,0.4){\epsfig{file=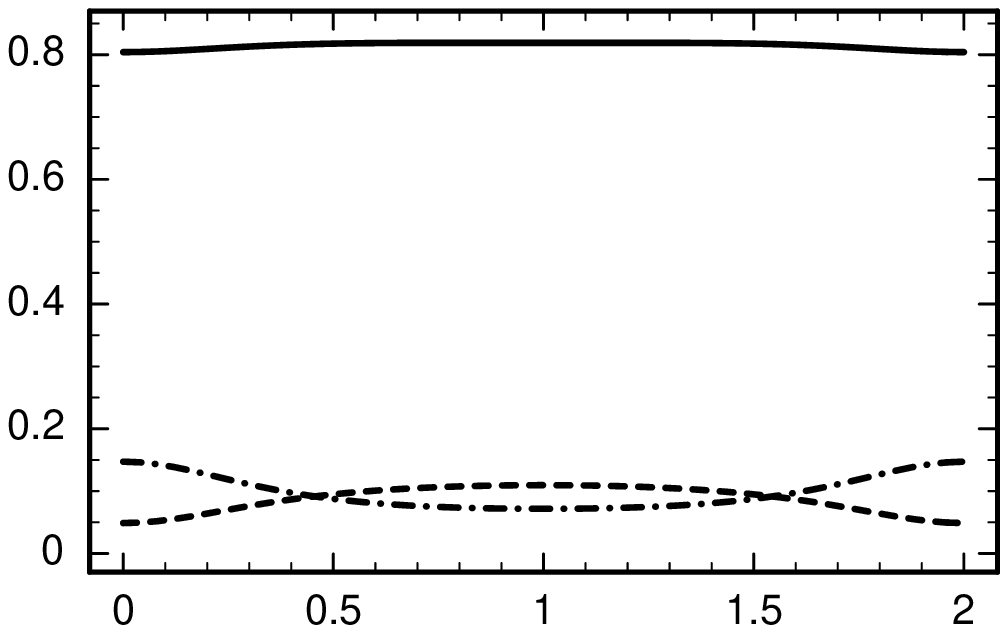,scale=.75}}
\put(8.4,5.2){$B$}
\put(14.8,0.1){$\varphi_{A_t}/\pi$}
\put(9,5.5){(f) $M_{\tilde{Q}}<M_{\tilde{U}}$, $|\mu|=350$~GeV}

\end{picture}
\caption{\label{brphiAt}(a), (d) Partial decay widths $\Gamma$ 
and (b), (c), (e), (f) branching ratios $B$ of the decays
$\tilde{t}_1 \to \tilde{\chi}^+_1 b$ (solid),
$\tilde{t}_1 \to \tilde{\chi}^+_2 b$ (dotted),
$\tilde{t}_1 \to \tilde{\chi}^0_1 t$ (dashed) and
$\tilde{t}_1 \to W^+ \tilde{b}_1$ (dashdotted)
for $\tan\beta = 6$, $M_2=300$~GeV, $|A_b|=|A_t|=800$~GeV, 
$\varphi_\mu=\pi$, $\varphi_\mathrm{U(1)}=\varphi_{A_b}=0$,
$m_{\tilde{t}_1}=350$~GeV, $m_{\tilde{t}_2}=700$~GeV,
$m_{\tilde{b}_1}=170$~GeV and $m_{H^\pm}=900$~GeV. 
In (a), (b) and (f) the decay $\ti t_1\to \ti\chi^+_2 b$ 
is kinematically forbidden.}

\end{figure}

We show in Fig.~\ref{brphiAt} the partial decay widths and branching ratios
for $\tilde{t}_1 \to \tilde{\chi}^+_1 b$,
$\tilde{t}_1 \to \tilde{\chi}^+_2 b$,
$\tilde{t}_1 \to \tilde{\chi}^0_1 t$ and 
$\tilde{t}_1 \to W^+ \tilde{b}_1$ as a function of $\varphi_{A_t}$ 
for the parameters
$\tan\beta = 6$, $M_2=300$~GeV, $|A_b|=|A_t|=800$~GeV,
$\varphi_\mu=\pi$, $\varphi_\mathrm{U(1)}=\varphi_{A_b}=0$,
$m_{H^\pm}=900$~GeV and two values of $|\mu| = 250$ and 350~GeV.
Figs.~\ref{brphiAt} (a) -- (c) (Figs.~\ref{brphiAt} (d) -- (f))
are for $M_{\tilde{Q}}>M_{\tilde{U}}$ ($M_{\tilde{Q}}<M_{\tilde{U}}$).
We first discuss Fig.~\ref{brphiAt} (a), (b)
for the case $M_{\tilde{Q}}>M_{\tilde{U}}$ and
$|\mu|=350$~GeV.
As can be seen in Fig.~\ref{brphiAt} (a), 
$\Gamma(\tilde{t}_1 \to \tilde{\chi}^+_1 b)$ and 
$\Gamma(\tilde{t}_1 \to \tilde{\chi}^0_1 t)$
show quite a significant
$\varphi_{A_t}$ dependence. The corresponding branching ratios are
shown in Fig.~\ref{brphiAt} (b). For $\varphi_{A_t}\approx 0$ and $2\pi$
the decay $\tilde{t}_1 \to \tilde{\chi}^0_1 t$ dominates, whereas for
$\varphi_{A_t}\approx \pi$ the decay 
$\tilde{t}_1 \to \tilde{\chi}^+_1 b$ has the
largest branching ratio.
This decay pattern can be explained in the
following way:
For $M_{\tilde{Q}}>M_{\tilde{U}}$ the $\tilde{t}_1$ is
$\tilde{t}_R$-like.
For $|\mu| > M_2$ and the parameters chosen, the chargino $(\ti\chi^\pm_1)$ is
$\tilde{W}^\pm$-like with $m_{\tilde{\chi}^\pm_1} = 279$~GeV, 
so that the decay $\ti t_1\to \ti\chi^+_1 b$ is suppressed by the
vanishing $\tilde{t}_R$-$b$-$\tilde{W}^+$ coupling and by small phase space. 
For the parameters chosen we have $|A_t| \gg |\mu|/\tan\beta$,
therefore $|M_{\tilde{t}RL}^2|$ and hence $\theta_{\tilde{t}}$ depend
only weakly on $\varphi_{A_t}$. However, we have $\varphi_{\tilde{t}}\approx
\varphi_{A_t}$ (see Eq.~(\ref{defphisquark})), 
therefore $\Gamma(\ti t_1\to \ti\chi^+_1 b)$ behaves like 
$(1 - \cos \varphi_{A_t})$:
the leading coupling term in this decay is 
$\ell^{\tilde{t}}_{11} = - e^{-i\varphi_{\tilde{t}}}
\cos\theta_{\tilde{t}} V_{11} + \sin\theta_{\tilde{t}} Y_t V_{12}$
(Eq.~(\ref{eq:ltij})), which consists of two terms of
comparable size, the phase $\varphi_{\tilde{t}} (\approx \varphi_{A_t})$
entering only in one of the two terms.
$\Gamma(\tilde{t}_1 \to \tilde{\chi}^+_1 b)$ is very small for
$\varphi_{A_t}=0$ and $2\pi$ because the two terms nearly cancel each other.
The $\varphi_{A_t}$ dependence of $\Gamma(\tilde{t}_1 \to \tilde{\chi}^0_1 t)$
is less pronounced compared to $\Gamma(\tilde{t}_1 \to \tilde{\chi}^+_1 b)$
due to a more complex coupling structure 
(Eqs.~(\ref{eq:abneut}) and (\ref{eq:abneut1})).
For this reason $B(\tilde{t}_1 \to \tilde{\chi}^0_1 t)$ 
dominates for $\varphi_{A_t} \lesssim 0.4 \pi$ and
$\varphi_{A_t} \gsim 1.6 \pi$, whereas
$B(\tilde{t}_1 \to \tilde{\chi}^+_1 b)$ is larger for 
$0.4\pi \lesssim \varphi_{A_t} \lesssim 1.6\pi$.
The branching ratio of $\tilde{t}_1 \to W^+ \tilde{b}_1$ is strongly suppressed
for this set of parameters with rather small $\tan\beta=6$ for which
$\tilde{b}_1$ is almost purely
$\tilde{b}_R$-like.
In Fig.~\ref{brphiAt} (c) we plot the branching ratios for 
$|\mu| = 250$~GeV.
In this case the lighter
chargino has a mass $m_{\tilde{\chi}^\pm_1} = 230$~GeV and a
significant higgsino component.
Hence the decay $\tilde{t}_1 \to \tilde{\chi}^+_1 b$
has a large phase space and large amplitude
(due to the large top Yukawa coupling $g Y_t$)
and dominates
independently of $\varphi_{A_t}$, resulting in a weak 
$\varphi_{A_t}$ dependence of the branching ratios.
For $|\mu| = 250$~GeV also the decay channel 
$\tilde{t}_1 \to \tilde{\chi}^+_2 b$ ($m_{\tilde{\chi}^\pm_2} = 336$~GeV) 
is open.

Figs.~\ref{brphiAt} (d) and (e) show the partial decay widths and
branching ratios of $\tilde{t}_1 \to \tilde{\chi}^+_1 b$,
$\tilde{\chi}^+_2 b$,
$\tilde{\chi}^0_1 t$ and 
$W^+ \tilde{b}_1$
against $\varphi_{A_t}$ for
$M_{\tilde{Q}}<M_{\tilde{U}}, |\mu|=250$~GeV and the other parameters as above.
In this case $\tilde{t}_1$ is $\tilde{t}_L$-like,
therefore for $|\mu|=250$~GeV (see Fig.~\ref{brphiAt} (d))
$\Gamma(\tilde{t}_1 \to \tilde{\chi}^+_1 b)$ 
is about three times as large as for
$M_{\tilde{Q}}>M_{\tilde{U}}$ and $|\mu|=350$~GeV (Fig.~\ref{brphiAt} (a)).
$\Gamma(\tilde{t}_1 \to \tilde{\chi}^+_1 b)$ behaves like
($1-\cos\varphi_{A_t}$), which
is again caused by an interplay of the two terms in the leading
coupling $\ell^{\tilde{t}}_{11}$ (Eq.~(\ref{eq:ltij})).
For $\varphi_{A_t} \approx 0$ the decay  
$\tilde{t}_1 \to \tilde{\chi}^+_1 b$ is suppressed
and the branching ratios of 
$\tilde{t}_1 \to \tilde{\chi}^+_2 b$,
$\tilde{t}_1 \to W^+ \tilde{b}_1$ and
$\tilde{t}_1 \to \tilde{\chi}^0_1 t$ reach
25\,\%, 22\,\% and 11\,\%, respectively (Fig.~\ref{brphiAt} (e)).
For $0.2\pi \lesssim \varphi_{A_t} \lesssim 1.8\pi$ 
the partial decay width and hence the branching ratio
of $\tilde{t}_1 \to \tilde{\chi}^+_1 b$ is clearly largest. 
$B(\tilde{t}_1 \to \tilde{\chi}^0_1 t)$ has values around 10\,\%.
$B(\tilde{t}_1 \to W^+ \tilde{b}_1)$ is rather small because
$\tilde{b}_1 \approx \tilde{b}_R$ in this case.
$\tilde{t}_1 \to \tilde{\chi}^+_2 b$ is suppressed by a 
small phase space.
In Fig.~\ref{brphiAt} (f) we show the corresponding branching ratios
for $M_{\ti Q}<M_{\ti U}$ and $|\mu| = 350$~GeV.
In this case the mixing in the bottom squark sector increases and
$B(\tilde{t}_1 \to W^+ \tilde{b}_1)$ reaches values around 10\,\%
even for $\varphi_{A_t}\approx\pi$.
The decay $\tilde{t}_1 \to \tilde{\chi}^+_1 b$ has the largest
branching ratio because $\tilde{t}_1$ is
$\tilde{t}_L$-like and $\tilde{\chi}^+_1$ is
almost $\tilde{W}^+$-like.
Hence in this scenario all branching ratios show a less pronounced
phase dependence.
In the scenarios of Fig.~\ref{brphiAt} we have calculated also the
$\varphi_\mathrm{U(1)}$ dependence of the partial decay
widths and branching ratios. 
By inspecting Eqs.~(\ref{eq:abneut})--(\ref{eq:htLk}) one can see that only
$\Gamma(\tilde{t}_1 \to \tilde{\chi}^0_1 t)$
could be sensitive to $\varphi_\mathrm{U(1)}$. However,
for $\tan\beta=6$ the $\varphi_\mathrm{U(1)}$ dependence
is already rather small.
This results in a weak $\varphi_\mathrm{U(1)}$
dependence of the branching ratios.

\begin{figure}[t]
\centering
\begin{picture}(16,5.7)

\put(0,.4){\epsfig{file=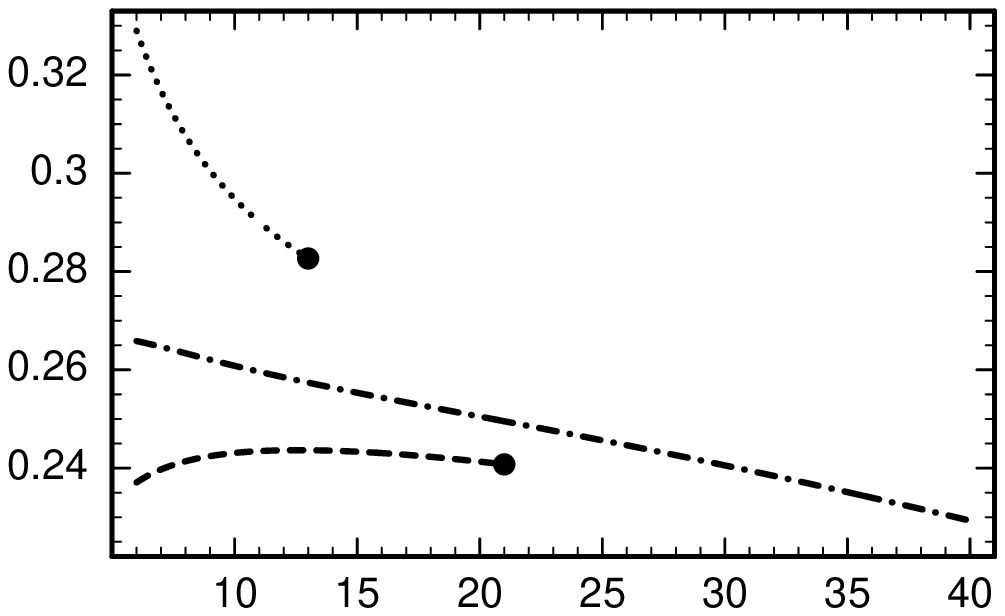,scale=.75}}    
\put(0.2,5.1){$B$}
\put(6.7,0.1){$\tan\beta$}
\put(5.5,5.3){(a) $\varphi_{A_t}=0$}

\put(8.4,.4){\epsfig{file=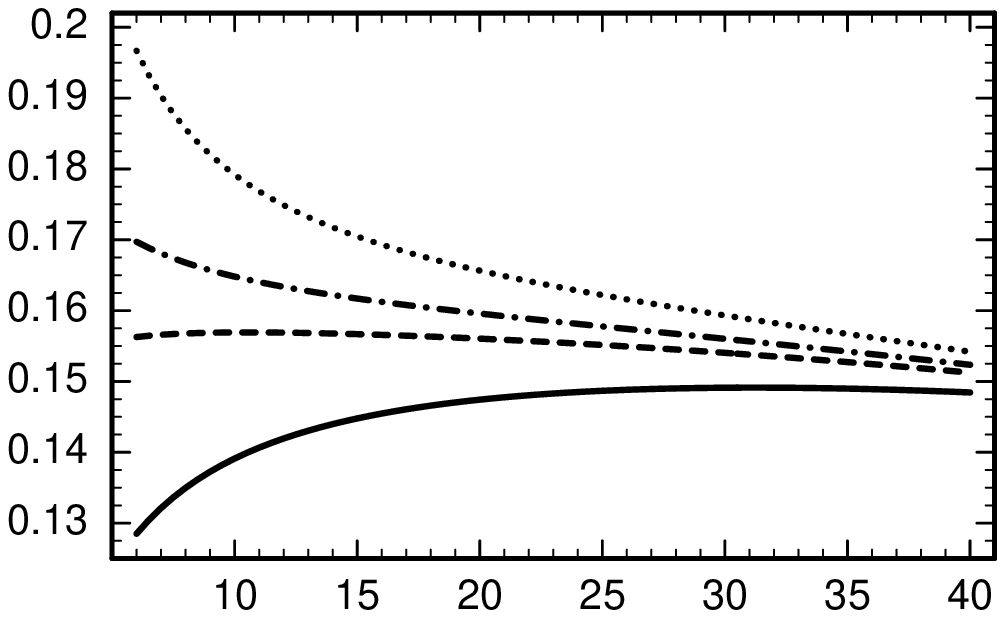,scale=.75}}    
\put(8.6,5.2){$B$}
\put(15.1,0.1){$\tan\beta$}
\put(13.9,5.3){(b) $\varphi_{A_t}=\pi$}

\end{picture}

\caption{\label{tanbeta}
Branching ratio $B(\tilde{t}_1 \to \tilde{\chi}^0_1 t)$ for 
$\varphi_\mu=0$ (solid), $\pi/2$ (dashed),
$5\pi/8$ (dashdotted) and $\pi$ (dotted) with
$\varphi_{A_t}=0$ (a) and $\pi$ (b),
$M_2=300$~GeV, $|\mu| = 300$~GeV, $|A_b|=|A_t|=600$~GeV, 
$\varphi_\mathrm{U(1)}=\varphi_{A_b}=0$,
$m_{\tilde{t}_1}=350$~GeV, $m_{\tilde{t}_2}=700$~GeV,
$m_{\tilde{b}_1}=170$~GeV and $m_{H^\pm}=500$~GeV, assuming
$M_{\tilde{Q}}>M_{\tilde{U}}$. 
In (a) the case $\varphi_\mu=0$ is excluded by the limit
$B(b\to s \gamma) < 4.5 \times 10^{-4}$, and the lines for $\varphi_\mu=\pi/2$
and $\varphi_\mu=\pi$ end in full circles beyond which 
$B(b\to s \gamma) > 4.5 \times 10^{-4}$ for ($\tan\beta\gtrsim 21$) and
$B(b\to s \gamma) < 2.0 \times 10^{-4}$ for ($\tan\beta\gtrsim 13$),
respectively.}

\end{figure}

In Fig.~\ref{tanbeta} we show the $\tan\beta$ dependence of
$B(\tilde{t}_1 \to \tilde{\chi}^0_1 t)$ for $M_2=300$~GeV, 
$|\mu| = 300$~GeV, $|A_b|=|A_t|=600$~GeV, 
$\varphi_\mathrm{U(1)}=\varphi_{A_b}=0$, $m_{H^\pm}=500$~GeV,
and $\varphi_\mu=0,\pi/2, 5\pi/8, \pi$ with (a) $\varphi_{A_t}=0$ and
(b) $\varphi_{A_t}=\pi$, assuming $M_{\tilde{Q}}>M_{\tilde{U}}$.
As can be seen this branching ratio
is insensitive to $\varphi_\mu$ for $\tan\beta\gtrsim
15$. This is mainly due to the $\mu/\tan\beta$ dependence of the
$\tilde{t}_L$-$\tilde{t}_R$ mixing term and
the insensitivity of the masses and mixing of $\ti\chi^0_i$ to 
$\varphi_\mu$ for large $\tan\beta$.
Two curves in Fig.~\ref{tanbeta}(a)
end in full circles beyond which the experimental
constraint from $B(b \to s \gamma)$ is violated: in case $\varphi_\mu=\pi/2$
($\varphi_\mu=\pi$), one has $B(b \to s \gamma) > 4.5 \times 10^{-4}$ 
($B(b \to s \gamma) < 2.0 \times 10^{-4}$) for $\tan\beta \gtrsim 21$
($\tan\beta \gtrsim 13$). The case $\varphi_\mu=0$ is completely
excluded for this set of parameters.
However, for $\varphi_{A_t}=\pi$ (Fig.~\ref{tanbeta} (b)) the
constraints from $B(b \to s \gamma)$ are always fulfilled.

We have also calculated the $\tan\beta$ dependence 
of the branching ratios of the $\tilde{t}_1$
decays for $M_{\tilde{Q}}<M_{\tilde{U}}$. 
$B(\tilde{t}_1 \to \tilde{\chi}^0_1 t)$ is smaller in this case.
Therefore the effect of the phase on the $\tan\beta$ dependence is also
smaller than in Fig.~\ref{tanbeta}.
Moreover, for $M_{\tilde{Q}}<M_{\tilde{U}}$ the situation is different
from that shown in Fig.~\ref{tanbeta}, because now for
$\varphi_{A_t}=0$ the whole $\tan\beta$ range is allowed, whereas for
$\varphi_{A_t}=\pi$ the constraints from $B(b \to s \gamma)$ limit the
$\tan\beta$ range.

\begin{figure}[t]
\centering
\begin{picture}(16,8.5)

\put(0,0.4){\epsfig{file=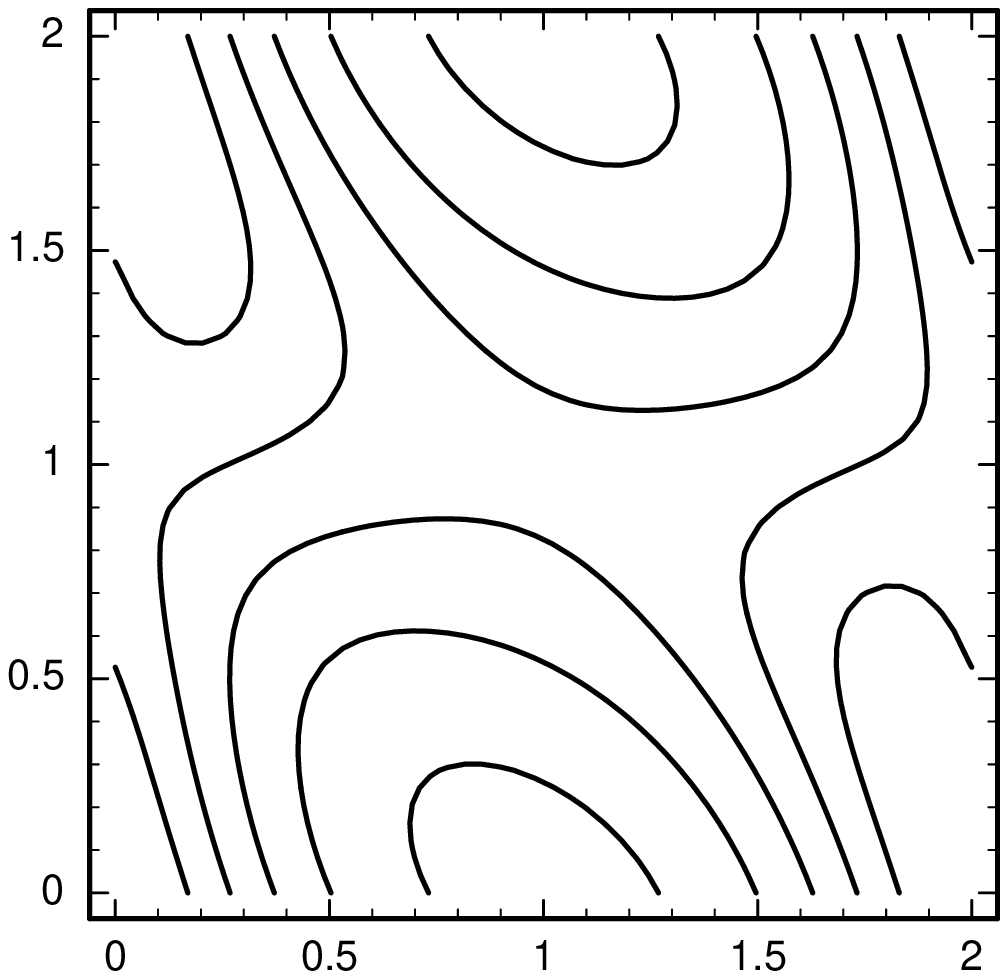,scale=.75}}
\put(0.1,8.1){$\varphi_\mu/\pi$}
\put(6.5,0.1){$\varphi_{A_t}/\pi$}
\put(3.8,8.1){(a)}
\put(0.85,1.2){\scriptsize 0.9}
\put(1.4,5){\scriptsize 0.9}
\put(2.1,4.3){\scriptsize 0.8}
\put(2.7,3.6){\scriptsize 0.7}
\put(2.8,2.8){\scriptsize 0.6}
\put(3.3,1.75){\scriptsize 0.5}
\put(4.6,6.8){\scriptsize 0.5}
\put(5.1,5.8){\scriptsize 0.6}
\put(5.2,4.95){\scriptsize 0.7}
\put(5.8,4.25){\scriptsize 0.8}
\put(6.5,3.55){\scriptsize 0.9}
\put(7,7.4){\scriptsize 0.9}

\put(8.4,0.4){\epsfig{file=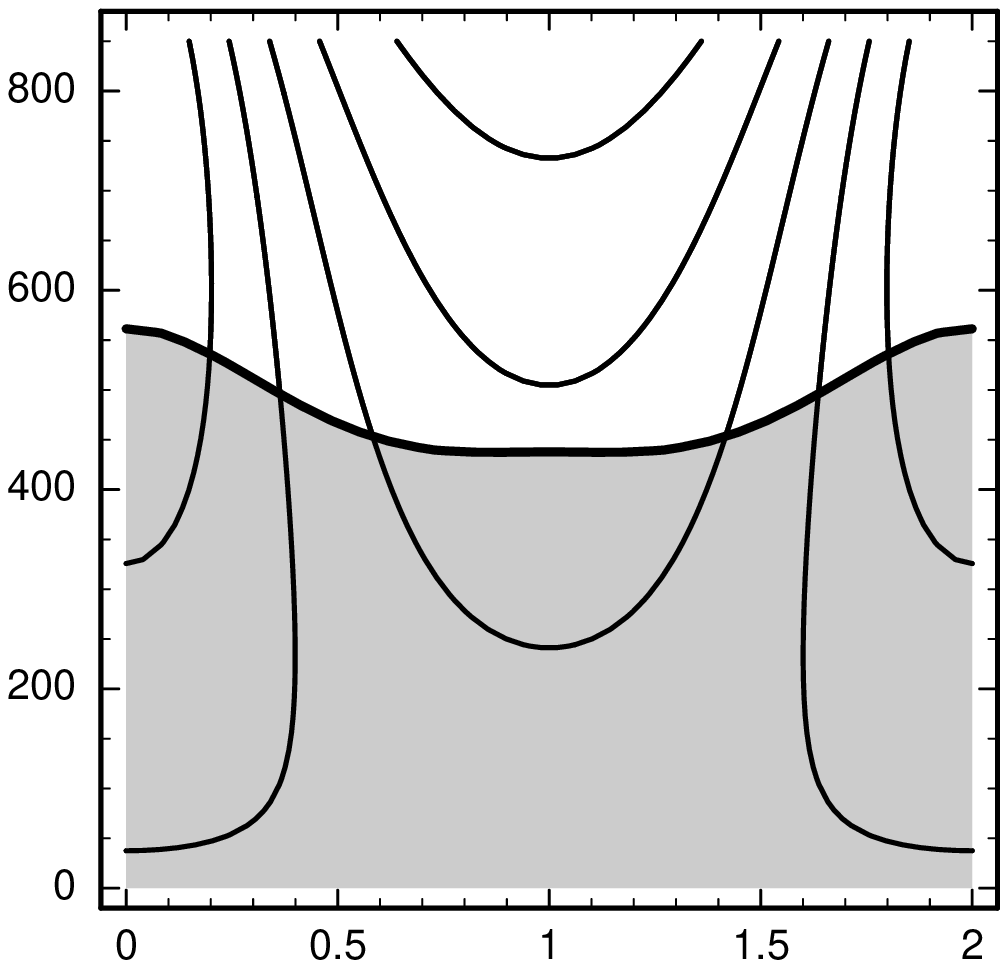,scale=.75}}
\put(8.4,8){$|A_t|/$GeV}
\put(14.9,0.1){$\varphi_{A_t}/\pi$}
\put(12.2,8.1){(b)}
\put(12.4,6.75){\scriptsize 0.5}
\put(12.4,5){\scriptsize 0.6}
\put(12.4,3){\scriptsize 0.7}
\put(10.65,1.7){\scriptsize 0.8}
\put(14.15,1.7){\scriptsize 0.8}
\put(9.6,3.35){\scriptsize 0.9}
\put(15.2,3.35){\scriptsize 0.9}

\end{picture}

\caption{\label{cpneut}Contours of $B(\tilde{t}_1 \to \tilde{\chi}^0_1 t)$ 
for $\tan\beta = 6$, $M_2=300$~GeV, $|\mu|=500$~GeV,
$\varphi_\mathrm{U(1)}=\varphi_{A_b}=0$,
$m_{\tilde{t}_1}=350$~GeV, $m_{\tilde{t}_2}=700$~GeV,
$m_{\tilde{b}_1}=170$~GeV, $m_{H^\pm}=600$~GeV,
with (a) $|A_t|=|A_b|=800$~GeV
and (b) $\varphi_\mu=0$, $|A_b|=|A_t|$, 
assuming $M_{\tilde{Q}}>M_{\tilde{U}}$.
The shaded area marks the region excluded by the Higgs search
at LEP (i.e.\ by the condition (ii)).}
\end{figure}

In Fig.~\ref{cpneut} (a) we show a contour plot for
$B(\tilde{t}_1 \to \tilde{\chi}^0_1 t)$ as a function of $\varphi_{A_t}$
and $\varphi_\mu$ for $\tan\beta = 6$, $M_2=300$~GeV,
$|\mu|=500$~GeV, $|A_t|=|A_b|=800$~GeV,
$\varphi_\mathrm{U(1)}=\varphi_{A_b}=0$ and $m_{H^\pm}=600$~GeV,
assuming $M_{\tilde{Q}}>M_{\tilde{U}}$. 
For the parameters chosen the $\varphi_{A_t}$ dependence is
stronger than the $\varphi_\mu$ dependence.
The reason is that these phase dependences are caused mainly by the
$\tilde{t}_L$ -$\tilde{t}_R$ mixing term (Eq.~(\ref{mLRterm})), where
the $\varphi_\mu$ dependence is suppressed by $\cot\beta$. The $\varphi_\mu$
dependence is somewhat more pronounced for $\varphi_{A_t} \approx \pi$
than for $\varphi_{A_t} \approx 0, 2\pi$.
In Fig.~\ref{cpneut} (b) we show the contour plot of 
$B(\tilde{t}_1 \to \tilde{\chi}^0_1 t)$ as a function of
$\varphi_{A_t}$ and $|A_t|$ for $\varphi_\mu = 0$ and
$|A_t|=|A_b|$. Clearly, the $\varphi_{A_t}$ dependence is stronger
for larger values of $|A_t|$.
For $M_{\tilde{Q}}<M_{\tilde{U}}$ we have obtained a similar behavior.
Note that the phase dependences of the decay branching ratios
of $\st_1 \to \tilde{\chi}^+_1 b$, $\st_1 \to \tilde{\chi}^+_2 b$ and
$\st_1 \to \tilde{\chi}^0_1 t$ analysed in Figs.~\ref{brphiAt},
\ref{tanbeta} and \ref{cpneut}
(where the decay $\st_1 \to \sb_1 H^+$ is kinematically forbidden)
would be present also for $\sb_1$ masses significantly larger than
170~GeV.

\begin{figure}[t]
\centering
\begin{picture}(8,8.5)

\put(0,0.4){\epsfig{file=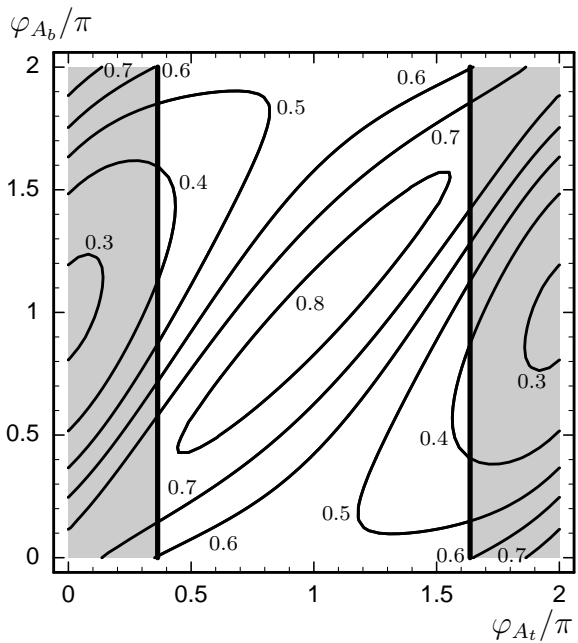,scale=.75}}
\put(0.1,8.1){$\varphi_{A_b}/\pi$}
\put(6.5,0.1){$\varphi_{A_t}/\pi$}
\put(1.1,5.2){\scriptsize 0.3}
\put(2.35,6){\scriptsize 0.4}
\put(3.65,7){\scriptsize 0.5}
\put(5.25,7.4){\scriptsize 0.6}
\put(5.7,6.6){\scriptsize 0.7}
\put(3.9,4.4){\scriptsize 0.8}
\put(2.2,1.95){\scriptsize 0.7}
\put(2.75,1.2){\scriptsize 0.6}
\put(4.25,1.6){\scriptsize 0.5}
\put(5.55,2.6){\scriptsize 0.4}
\put(6.85,3.35){\scriptsize 0.3}
\put(2.12,7.5){\scriptsize 0.6}
\put(1.35,7.5){\scriptsize 0.7}
\put(5.77,1.05){\scriptsize 0.6}
\put(6.58,1.05){\scriptsize 0.7}

\end{picture}

\caption{\label{cpcharg}Contours of $B(\tilde{t}_1 \to \tilde{\chi}^+_1 b)$ 
as a function of $\varphi_{A_t}$ and $\varphi_{A_b}$ 
for $\tan\beta = 30$, $M_2=300$~GeV, $|\mu|=300$~GeV,
$|A_b|=|A_t|=600$~GeV, $\varphi_\mu=\pi$, $\varphi_\mathrm{U(1)}=0$,
$m_{\tilde{t}_1}=350$~GeV, $m_{\tilde{t}_2}=700$~GeV,
$m_{\tilde{b}_1}=170$~GeV and $m_{H^\pm}=160$~GeV, assuming
$M_{\tilde{Q}}>M_{\tilde{U}}$.
The shaded areas are excluded by the experimental limit
$B(b\to s \gamma) > 2.0 \times 10^{-4}$.}

\end{figure}

In Fig.~\ref{cpcharg} we show the contour plot for
$B(\tilde{t}_1 \to \tilde{\chi}^+_1 b)$ as a function of
$\varphi_{A_t}$ and $\varphi_{A_b}$ for
$\tan\beta = 30$, $M_2=300$~GeV, $|\mu|=300$~GeV,
$|A_b|=|A_t|=600$~GeV, $\varphi_\mu=\pi$, $\varphi_\mathrm{U(1)}=0$ and
$m_{H^\pm}=160$~GeV, assuming $M_{\tilde{Q}}>M_{\tilde{U}}$.
As can be seen, there is a remarkable correlation between 
$\varphi_{A_t}$ and $\varphi_{A_b}$, which turns out to be relatively
independent of $\varphi_\mu$.
The $\varphi_{A_t}$-$\varphi_{A_b}$ correlation
can be explained by the behavior of the 
partial decay width $\Gamma(\tilde{t}_1 \to H^+ \tilde{b}_1)$, which
influences all decay branching 
ratios. As $\ti b_1\sim\ti b_R$ in this case, the 
relevant coupling for $\tilde{t}_1 \to H^+ \tilde{b}_1$ is
$C^H_{\ti b_1\ti t_1}\sim (\mathcal{R}^{\tilde{t}} G)^{\ast}_{12}$ (see
Eq.~(\ref{eq:G})). $\mathcal{R}^{\tilde{t}}$ depends on
$\varphi_{A_t}$ via $\tilde{t}_L$-$\tilde{t}_R$ mixing, whereas $G$
depends on $\varphi_{A_b}$ via the coupling term
$m_b\,(A_b^{\ast}\tan\b + \mu)$. 
As $\varphi_{\ti t}\approx\varphi_{A_t}$ in this case, we have
$(\mathcal{R}^{\tilde{t}} G)_{12} \approx e^{i(\varphi_{A_t} -
  \varphi_{A_b})} \cos\theta_{\tilde{t}} \cdot m_b |A_b| \tan\b +
  \sin\theta_{\tilde{t}} \cdot 2 m_t m_b/\sin 2\b$
which clearly shows the correlation between $\varphi_{A_t}$ and
$\varphi_{A_b}$ apart from the much weaker $\varphi_{A_t}$ dependence of 
$\theta_{\tilde{t}}$.
Note that here the small value for the $\sb_1$ mass
($m_{\sb_1}=170$~GeV) is important: for a larger $\sb_1$ mass the
decay $\st_1 \to \sb_1 H^+$ would not 
be allowed kinematically and hence the $\varphi_{A_b}$ dependence shown
in Fig.~\ref{cpcharg} would disappear.

For $M_{\tilde{Q}}<M_{\tilde{U}}$ the decay 
$\tilde{t}_1 \to H^+ \tilde{b}_1$ dominates for all $\varphi_{A_t}$
and $\varphi_{A_b}$ resulting in a weaker phase dependence of all
branching ratios. Hence also the correlation between $\varphi_{A_t}$
and $\varphi_{A_b}$ in $B(\tilde{t}_1 \to \tilde{\chi}^+_1 b)$ is less
pronounced.
However, in the scenario of Fig.~\ref{cpcharg} one has
$B(b\to s\gamma) > 4.7 \times 10^{-4}$ for $M_{\tilde{Q}}<M_{\tilde{U}}$.

\begin{figure}[t]
\centering
\begin{picture}(16,5.7)

\put(0,0.4){\epsfig{file=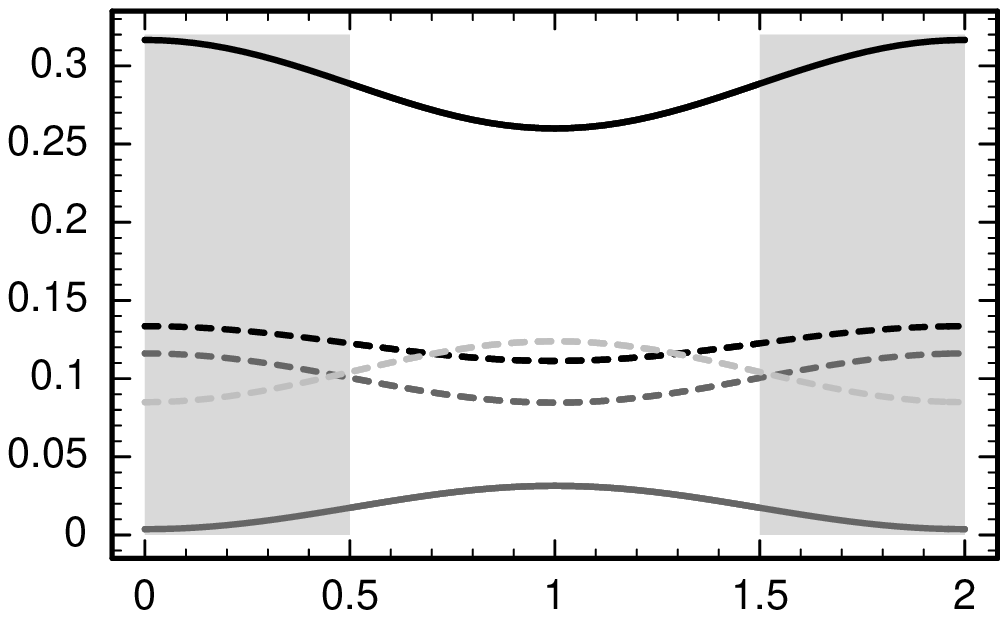,scale=.75}}    
\put(0.3,5){$B$}
\put(6.5,0.1){$\varphi_{A_t}/\pi$}
\put(3.95,5.3){(a)}

\put(8.4,0.4){\epsfig{file=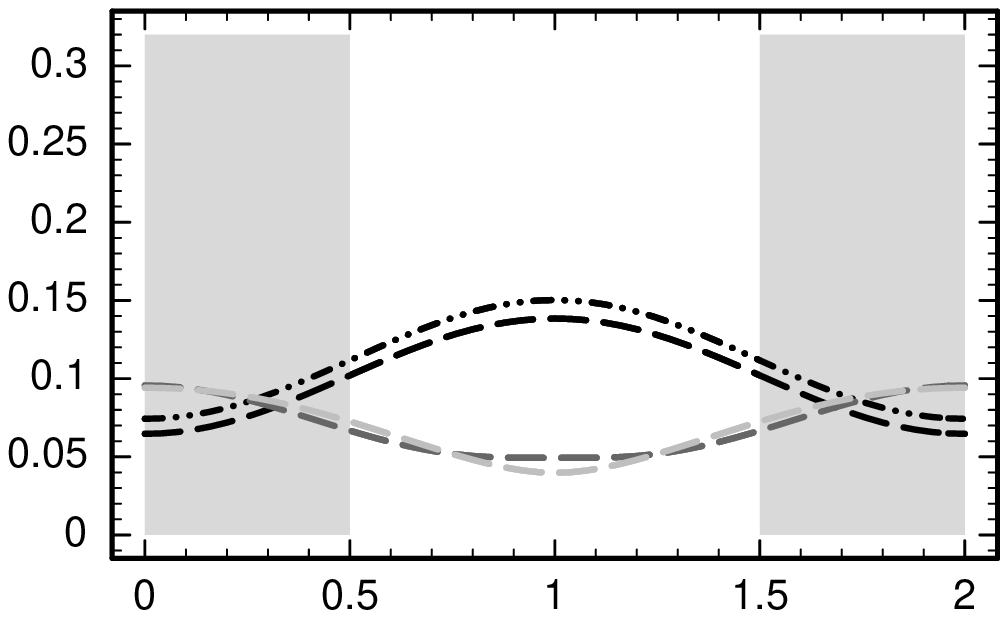,scale=.75}}    
\put(8.6,5){$B$}
\put(14.9,0.1){$\varphi_{A_t}/\pi$}
\put(12.35,5.3){(b)}

\end{picture}

\caption{\label{brphiAtSt2} $\varphi_{A_t}$ dependence of
branching ratios of the decays (a)
$\tilde{t}_2 \to \tilde{\chi}^+_{1/2} b$ (solid, black/gray),
$\tilde{t}_2 \to \tilde{\chi}^0_{2/3/4} t$ (dashed, black/gray/light gray)
and (b)
$\tilde{t}_2 \to Z \tilde{t}_1$ (dashdotdotted),
$\tilde{t}_2 \to H_{1/2/3} \tilde{t}_1$ (long dashed, black/gray/light
gray)
for $\tan\beta = 6$,
$M_2=300$~GeV, $|\mu| = 500$~GeV, $|A_b|=|A_t|=500$~GeV,
$\varphi_\mu=\varphi_\mathrm{U(1)}=\varphi_{A_b}=0$,
$m_{\tilde{t}_1}=350$~GeV, $m_{\tilde{t}_2}=800$~GeV,
$m_{\tilde{b}_1}=170$~GeV and $m_{H^\pm}=350$~GeV, assuming
$M_{\tilde{Q}}>M_{\tilde{U}}$. Only the decay modes with $B \gtrsim
1\,\%$ are shown.
The shaded areas mark the region excluded by the
experimental limit $B(b\to s \gamma) < 4.5 \times 10^{-4}$.}

\end{figure}

For the heavier top squark $\tilde{t}_2$ more decay channels are open. 
Besides the fermionic decay modes 
$\tilde{t}_2 \to \tilde{\chi}^+_j b$, 
$\tilde{\chi}^0_i t$ ($j=1,2$; $i=1,\dots,4$) there are also
the bosonic decay modes
$\tilde{t}_2 \to W^+ \ti b_j, H^+ \ti b_j, Z \tilde{t}_1, H_i \tilde{t}_1$
($j=1,2$; $i=1,2,3$). 
In Fig.~\ref{brphiAtSt2} (a) we show the branching ratios for 
$\tilde{t}_2 \to \tilde{\chi}^+_{1,2} b$ and
$\tilde{t}_2 \to \tilde{\chi}^0_{2,3,4} t$
as a function of $\varphi_{A_t}$ for  $\tan\beta = 6$,
$M_2=300$~GeV, $|\mu| = 500$~GeV, $|A_b|=|A_t|=500$~GeV,
$\varphi_\mu=\varphi_\mathrm{U(1)}=\varphi_{A_b}=0$, 
$m_{\tilde{t}_1}=350$~GeV, $m_{\tilde{t}_2}=800$~GeV,
$m_{\tilde{b}_1}=170$~GeV and $m_{H^\pm}=350$~GeV,
assuming $M_{\tilde{Q}}>M_{\tilde{U}}$.
The $\varphi_{A_t}$ dependence of 
$B(\tilde{t}_2 \to \tilde{\chi}^+_{1,2} b)$ is again due to a direct phase
effect, because the leading coupling $\ell^{\tilde{t}}_{2j}$, $j=1,2$
(Eq.~(\ref{eq:ltij}))
consists of two terms, with the phase $\varphi_{\tilde{t}} (\approx
\varphi_{A_t})$ entering only the
factor ${\mathcal{R}^{\tilde{t}}_{22}}^{\!\!\!\ast}$
in the second term. Therefore, the shape of 
$B(\tilde{t}_2 \to \tilde{\chi}^+_{1,2} b)$ is like 
($1\pm\cos\varphi_{A_t}$).
Also the phase dependence of the branching ratios into neutralinos
is mainly due to a direct phase effect.
In $\Gamma(\tilde{t}_2 \to \tilde{\chi}^0_i t)$, $i=2,3,4$
the phase $\varphi_{\tilde{t}} (\approx \varphi_{A_t})$ enters into the
second term of the couplings $a^{\tilde{t}}_{2i}$ and 
$b^{\tilde{t}}_{2i}$ (see
Eq.~(\ref{eq:abneut})) via  ${\mathcal{R}^{\tilde{t}}_{22}}^{\!\!\!\ast}$.
For $\Gamma(\tilde{t}_2 \to \tilde{\chi}^0_2 t)$ the coupling
$a^{\tilde{t}}_{22}$ dominates and the size of its second term is smaller
than 10\,\% of its first term.
Hence the $|a^{\tilde{t}}_{22}|^2$ term in the
width of Eq.~(\ref{eq:gamtN}) creates its weak phase dependence like
($10+\cos\varphi_{A_t}$).
However, for $\Gamma(\tilde{t}_2 \to \tilde{\chi}^0_3 t)$ the 
mixing phase enters
mainly into the second term of the partial width via
$\mathrm{Re}(a_{23}^{\tilde{t}*} b_{23}^{\tilde{t}})\sim
\mathrm{Re}({\mathcal{R}^{\tilde{t}}_{22}}
{\mathcal{R}^{\tilde{t}}_{21}}^{\!\!\!\ast})\sim
\cos\varphi_{\ti t}\sim \cos\varphi_{A_t}$, 
resulting in a shape like ($1+\cos\varphi_{A_t}$).
For $\Gamma(\tilde{t}_2 \to \tilde{\chi}^0_4 t)$ the two terms in
$a^{\tilde{t}}_{24}$ have comparable size resulting in a strong $\varphi_{A_t}$
dependence of the terms $|a^{\tilde{t}}_{24}|^2$ and 
$\mathrm{Re}(a_{24}^{\tilde{t}*} b_{24}^{\tilde{t}})$ in the partial
width which eventually causes the branching ratio to behave like
($1-\cos\varphi_{A_t}$).

In Fig.~\ref{brphiAtSt2} (b) we show the branching ratios for the
bosonic decays $\tilde{t}_2 \to Z \tilde{t}_1$ and 
$\tilde{t}_2 \to H_i \tilde{t}_1$ $(i=1,2,3)$ for the same parameter values
as above.
The shape of $B(\tilde{t}_2 \to Z \tilde{t}_1)$ is
like $(1-\cos\varphi_{A_t})$ , which is
solely due to the factor $|\sin 2 \theta_{\tilde{t}}|^2$ 
(see Eq.~(\ref{eq:cij})). 
Quite generally, the
phase dependence of 
$\Gamma(\ti t_2\to H_k\ti t_1)$ 
is the result of a complicated
interplay among the phase dependences of the $H_k$ 
masses, the top squark mixing matrix
elements ${\mathcal{R}^{\tilde{t}}_{ij}}$, 
the neutral Higgs mixing matrix elements $O_{ij}$
and the direct top squark-Higgs couplings of $\ti t_L\ti t_R\phi_{1,2}$
and $\ti t_L\ti t_R a$.
In the present example the $\varphi_{A_t}$ dependence of the partial
widths $\Gamma(\tilde{t}_2 \to H_{1,2,3} \tilde{t}_1)$ is mainly due
to the $\varphi_{A_t}$ dependence of the factors
$\mathcal{R}^{\tilde{t}}$ and $C(\st_{L}^{\dagger} H_i \st_{R})$ in 
Eqs.~(\ref{eq:Hijcoup}) -- (\ref{eq:HRLcoupt}), whereas the
$\varphi_{A_t}$ dependence of the $O_{ij}$ is less pronounced in
this case\footnote{
For completeness we remark that the effect of the phase
dependence of $\tilde t_i-\tilde t_j-H_k$ couplings also
shows up in processes like
$e^+ e^- \to \tilde t_1 \bar{\tilde t}_1 H_1$
\cite{Bae:2000fn}.
}.

We have also calculated the branching ratios of the $\tilde{t}_2$
decays for $M_{\tilde{Q}} < M_{\tilde{U}}$.
In this case no constraints on the $\varphi_{A_t}$ range
from the $B(b\to s \gamma)$ data arise in the given
scenario.
The $\varphi_{A_t}$ dependence of $B(\tilde{t}_2 \to Z \tilde{t}_1)$ and 
$B(\tilde{t}_2 \to H_{1,2,3} \tilde{t}_1)$ is very similar to that
shown in Fig.~\ref{brphiAtSt2} (b). The leading branching ratios are
now $B(\tilde{t}_2 \to \tilde{\chi}^+_2 b)$,
$B(\tilde{t}_2 \to H^+ \tilde{b}_2)$ and 
$B(\tilde{t}_2 \to W^+ \tilde{b}_2)$ with the values
17\,\%, 15\,\% and 13\,\% for $\varphi_{A_t} = 0,2\pi$ and
18\,\%, 7\,\% and 24\,\% for $\varphi_{A_t} = \pi$, respectively.

Furthermore we have calculated the $\varphi_\mathrm{U(1)}$ dependence of the
branching ratios of the $\tilde{t}_2$ for the scenario of 
Fig.~\ref{brphiAtSt2}.
It turns out to be very weak because
$\Gamma(\tilde{t}_2 \to \tilde{\chi}^0_1 t)$ 
(with $\tilde{\chi}^0_1\sim \ti B$) is suppressed in this scenario.
The $\varphi_{U(1)}$ dependence stems only from that of
$\Gamma(\ti t_2\to\ti\chi^0_i t)$, $i=2,3,4$.
$\tilde{\chi}^0_2$ and $\tilde{\chi}^0_{3,4}$ are wino- and higgsino-dominated,
respectively. 
Hence the masses $m_{\ti\chi^0_{2,3,4}}$ and mixings $N_{ij}$ ($i=2,3,4$) of 
$\ti\chi^0_{2,3,4}$ are rather insensitive to
the bino phase $\varphi_\mathrm{U(1)}$.

\subsection{Bottom squark decays}

In the discussion of $\tilde{b}_{1,2}$ decays
we fix $\tan\beta = 30$ because for small $\tan\beta$ the bottom squark mixing
is too small to be phenomenologically interesting. 
We fix the other parameters as 
$m_{\tilde{b}_1}=350$~GeV, $m_{\tilde{b}_2}=700$~GeV,
$m_{\tilde{t}_1}=170$~GeV, $m_{H^\pm}=150$~GeV and
$M_2=200$~GeV.
We have chosen a relatively small value for the $\st_1$ mass
to allow for the decay $\sb_1 \to H^- \st_1$ which has a rather
strong dependence on $\varphi_{A_b}$.

\begin{figure}[t]
\centering
\begin{picture}(16,6)

\put(0,0.4){\epsfig{file=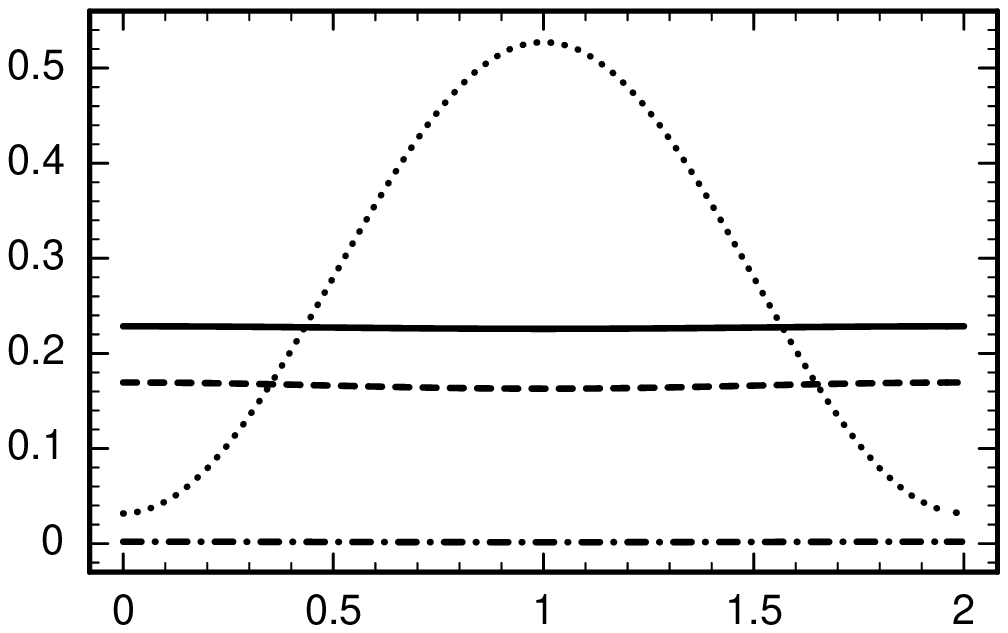,scale=.75}}
\put(0.1,5.35){$\Gamma/$GeV}
\put(6.4,0.1){$\varphi_{A_b}/\pi$}
\put(3.9,5.4){(a)}

\put(8.4,0.4){\epsfig{file=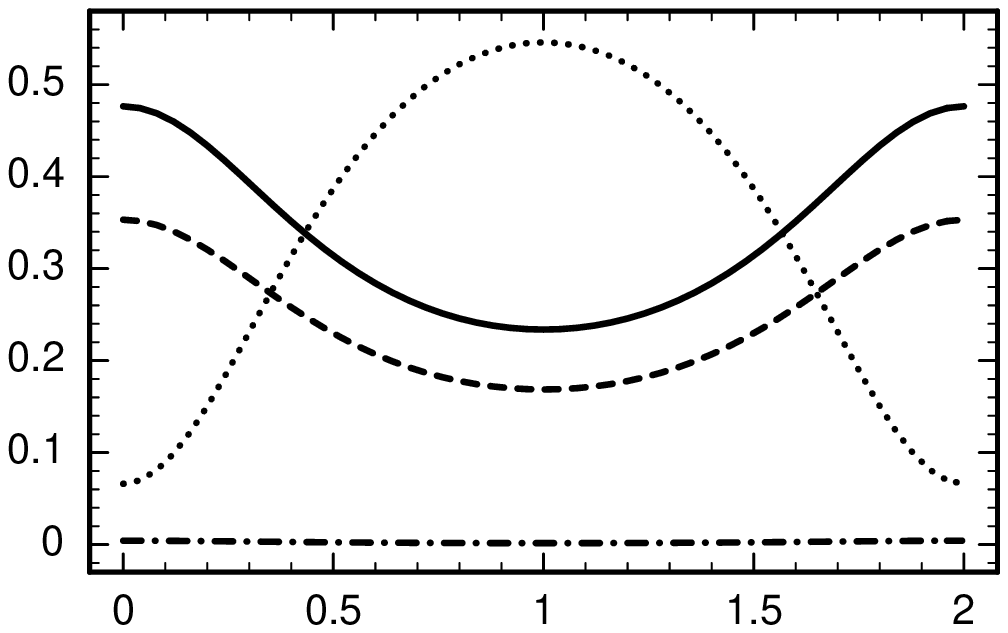,scale=.75}}
\put(8.5,5.2){$B$}
\put(14.8,0.1){$\varphi_{A_b}/\pi$}
\put(12.3,5.4){(b)}

\end{picture}

\caption{\label{sbottomphiAb} $\varphi_{A_b}$ dependences of
(a) partial widths and (b) branching
ratios of the decays
$\tilde{b}_1 \to \tilde{\chi}^0_1 b$ (solid),
$\tilde{b}_1 \to \tilde{\chi}^0_2 b$ (dashed),
$\tilde{b}_1 \to H^- \tilde{t}_1$ (dotted) and
$\tilde{b}_1 \to W^- \tilde{t}_1$ (dashdotted)
for $\tan\beta = 30$, $M_2=200$~GeV, $|\mu| = 300$~GeV, $|A_b|=|A_t|=600$~GeV,
$\varphi_\mu=\pi$, $\varphi_{A_t}=\varphi_\mathrm{U(1)}=0$,
$m_{\tilde{b}_1}=350$~GeV, $m_{\tilde{b}_2}=700$~GeV,
$m_{\tilde{t}_1}=170$~GeV and $m_{H^\pm}=150$~GeV, assuming
$M_{\tilde{Q}}>M_{\tilde{D}}$.}

\end{figure}

In Fig.~\ref{sbottomphiAb} we show the partial decay widths and
the branching ratios of $\tilde{b}_1 \to \tilde{\chi}^0_{1,2} b$,
$H^- \tilde{t}_1$, $W^- \tilde{t}_1$
as a function of $\varphi_{A_b}$ for $|\mu| = 300$~GeV,
$|A_b|=|A_t|=600$~GeV,
$\varphi_\mu=\pi$ and $\varphi_{A_t}=\varphi_\mathrm{U(1)}=0$, 
assuming $M_{\tilde{Q}}>M_{\tilde{D}}$.
In the region $0.5 \pi < \varphi_{A_b} < 1.5 \pi$ the decay
$\tilde{b}_1 \to H^- \tilde{t}_1$ dominates.
The $\varphi_{A_b}$ dependence of 
$\Gamma(\tilde{b}_1 \to H^- \tilde{t}_1)$ is due to the term
$m_b |A_b| e^{-i\varphi_{A_b}} \tan\beta$ in Eq.~(\ref{eq:G1}).
The partial decay widths $\Gamma(\tilde{b}_1 \to \tilde{\chi}^0_{1,2} b)$ are
almost $\varphi_{A_b}$ independent because the $\varphi_{A_b}$
dependence of the $\ti b$-mixing matrix $\mathcal{R}^{\tilde{b}}$
nearly vanishes for $\tan\beta=30$.
Hence the $\varphi_{A_b}$ dependence of the branching ratios 
$B(\tilde{b}_1 \to \tilde{\chi}^0_{1,2} b)$ is caused by that
of the total decay width.
$\Gamma(\tilde{b}_1 \to W^- \tilde{t}_1)$ is suppressed because 
$\tilde{b}_1\sim \tilde{b}_R$ and $\tilde{t}_1\sim \tilde{t}_R$ 
in this scenario (since also $M_{\tilde{Q}} > M_{\tilde{U}}$).
For the scenario of Fig.~\ref{sbottomphiAb} the case
$M_{\tilde{Q}} < M_{\tilde{D}}$ is excluded by the experimental lower limit
$B(b\to s \gamma) > 2.0 \times 10^{-4}$.

The $\varphi_\mathrm{U(1)}$ dependence of the partial decay widths 
and branching ratios is very weak in the scenario of
Fig.~\ref{sbottomphiAb}. $\varphi_\mathrm{U(1)}$ enters only into
$\Gamma(\tilde{b}_1 \to \tilde{\chi}^0_{k} b)$ ($k=1,2$) which are nearly
independent of $\varphi_\mathrm{U(1)}$ because 
$\tilde{b}_1 \sim \tilde{b}_R$ and hence mainly 
$h^b_{Rk}$ in $a^{\tilde{b}}_{1k}$ and 
$f^b_{Rk}$ in $b^{\tilde{b}}_{1k}$
contribute (see Eqs.~(\ref{eq:abneut}) -- (\ref{eq:hbLk})).
Then the phase of $N_{k1}$, 
which strongly depends on $\varphi_\mathrm{U(1)}$, almost drops out 
in Eq.~(\ref{eq:gamtN}). Furthermore, the masses 
$m_{\tilde{\chi}^0_i}$ and mixing
matrix elements $N_{ij}$ of the $\tilde{\chi}^0_i$-sector are insensitive
to $\varphi_\mathrm{U(1)}$ for large $\tan\beta$.

\begin{figure}[t]
\centering
\begin{picture}(16,8.5)

\put(0,0.4){\epsfig{file=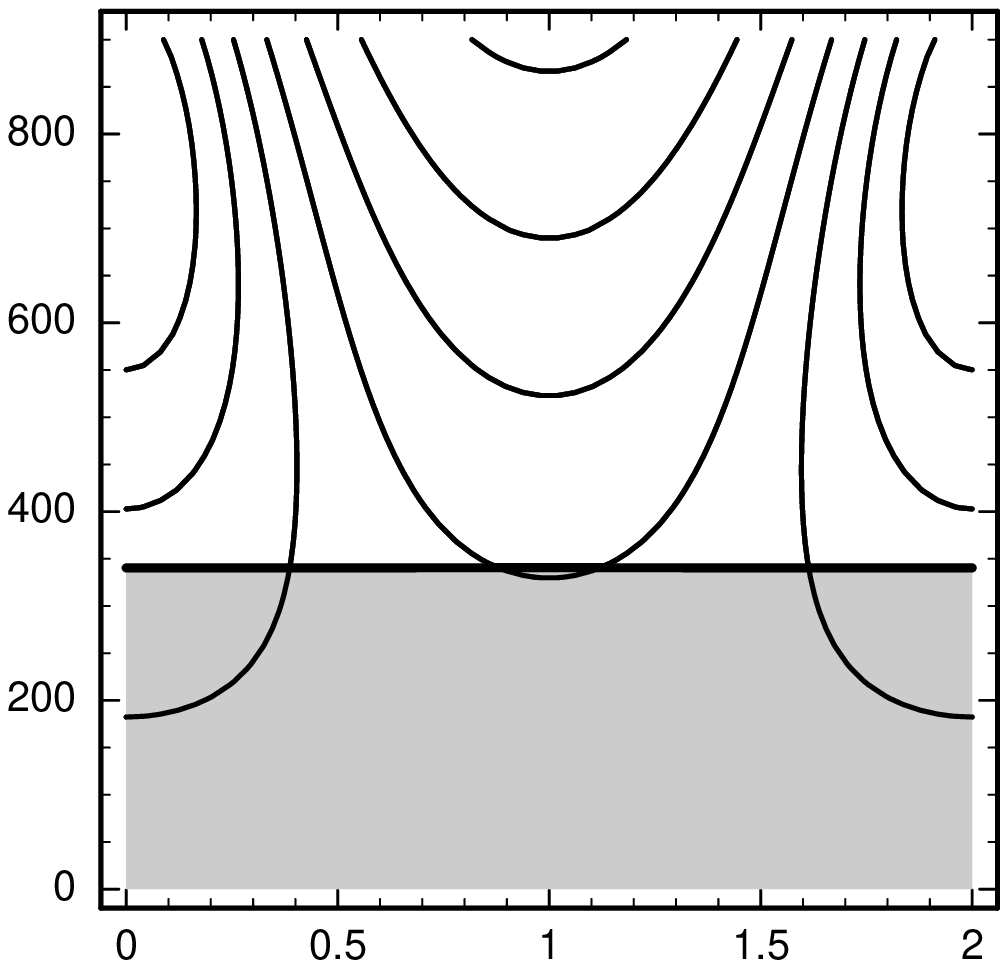,scale=.76}}
\put(0.1,8.1){$|A_b|/$GeV}
\put(6.5,0.1){$\varphi_{A_b}/\pi$}
\put(4,8.1){(a)}
\put(0.95,4.85){\tiny 0.1}
\put(0.95,3.78){\tiny 0.2}
\put(0.95,2.18){\tiny 0.3}
\put(7.2,2.18){\tiny 0.3}
\put(7.2,3.78){\tiny 0.2}
\put(7.2,4.85){\tiny 0.1}
\put(4.05,3.6){\tiny 0.4}
\put(4.05,4.95){\tiny 0.5}
\put(4.05,6.18){\tiny 0.6}
\put(4.05,7.45){\tiny 0.7}

\put(8.4,0.4){\epsfig{file=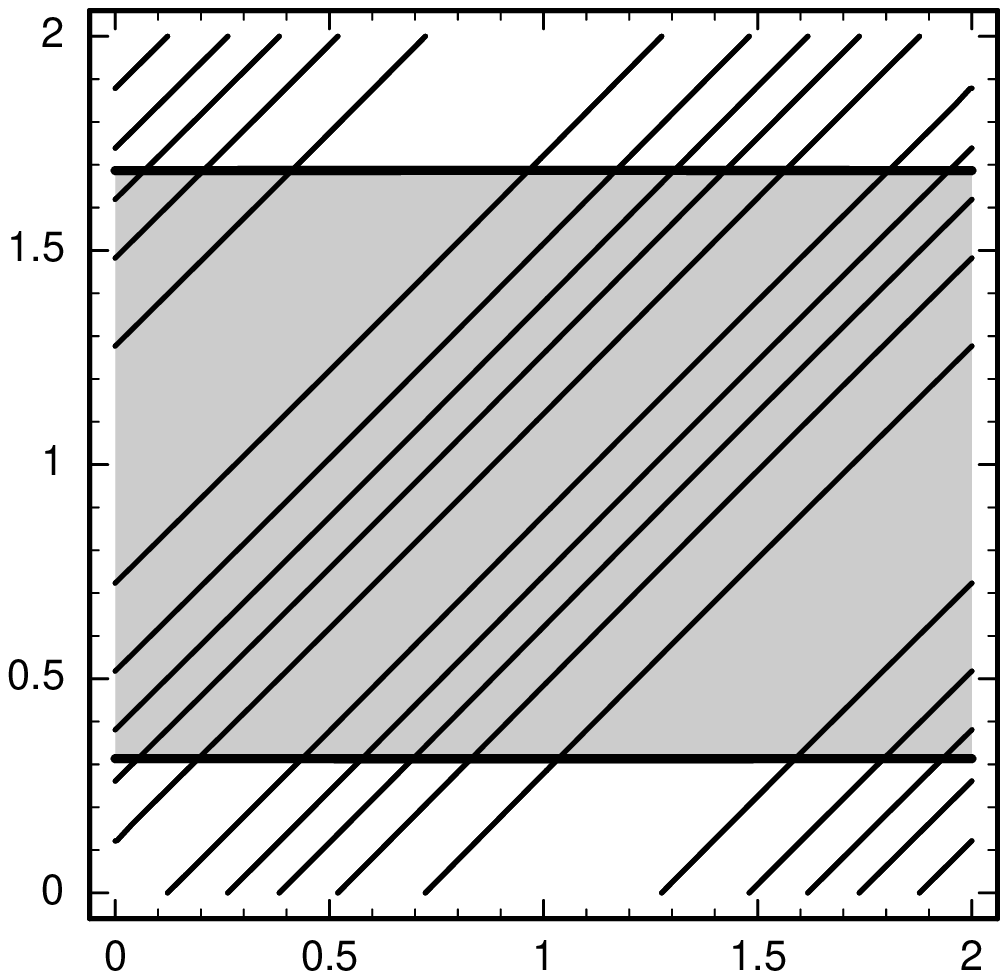,scale=.75}}
\put(8.4,8.1){$\varphi_{A_t}/\pi$}
\put(14.9,0.1){$\varphi_{A_b}/\pi$}
\put(12.3,8.1){(b)}
\put(9.28,7.6){\tiny 0.1}
\put(9.73,7.6){\tiny 0.2}
\put(10.55,7.6){\tiny 0.3}
\put(11,7.6){\tiny 0.4}
\put(11.7,7.6){\tiny 0.5}
\put(13,7.6){\tiny 0.5}
\put(13.7,7.6){\tiny 0.4}
\put(14.15,7.6){\tiny 0.3}
\put(14.97,7.6){\tiny 0.2}
\put(15.45,7.6){\tiny 0.1}

\put(9.35,1.03){\tiny 0.1}
\put(9.77,1.03){\tiny 0.2}
\put(10.58,1.03){\tiny 0.3}
\put(11.05,1.03){\tiny 0.4}
\put(11.75,1.03){\tiny 0.5}
\put(13.05,1.03){\tiny 0.5}
\put(13.75,1.03){\tiny 0.4}
\put(14.2,1.03){\tiny 0.3}
\put(15,1.03){\tiny 0.2}
\put(15.48,1.03){\tiny 0.1}

\end{picture}

\caption{\label{cpsbottom}Contours of $B(\tilde{b}_1 \to H^- \tilde{t}_1)$
for $\tan\beta = 30$, $M_2=200$~GeV, $|\mu|=300$~GeV,
$\varphi_\mu=\pi$, $\varphi_\mathrm{U(1)}=0$,
$m_{\tilde{b}_1}=350$~GeV, $m_{\tilde{b}_2}=700$~GeV,
$m_{\tilde{t}_1}=170$~GeV, $m_{H^\pm}=150$~GeV and
(a)  $|A_b|=|A_t|$, $\varphi_{A_t}=0$
and (b) $|A_b|=|A_t|=600$~GeV,
assuming $M_{\tilde{Q}}>M_{\tilde{D}}$.
The shaded areas in (a) and (b) mark the regions excluded by the
Higgs search at LEP (i.e.\ by the condition (ii)) and by the
experimental limit $B(b\to s \gamma) > 2.0 \times 10^{-4}$,
respectively.}

\end{figure}

For large $\tan\beta$ one expects also a significant $|A_b|$
dependence of 
$\Gamma(\tilde{b}_1 \to H^- \tilde{t}_1)$ (see Eq.~(\ref{eq:G1})).
This can be seen in
Fig.~\ref{cpsbottom} (a) where we show the contour plot of
$B(\tilde{b}_1 \to H^- \tilde{t}_1)$ as a function of
$|A_b|$ and $\varphi_{A_b}$ for $|\mu|=300$~GeV,
$\varphi_\mu=\pi$, $\varphi_{A_t}=\varphi_\mathrm{U(1)}=0$ and 
$|A_t| =  |A_b|$, assuming $M_{\tilde{Q}}>M_{\tilde{D}}$. The $\varphi_{A_b}$ 
dependence is stronger for larger
values of $|A_b|$. Although Fig.~\ref{cpsbottom} (a) is similar to
Fig.~\ref{cpneut} (b), the $|A_b|$ and $\varphi_{A_b}$ dependence in
Fig.~\ref{cpsbottom} (a) is now caused by the coupling
$m_b\,(A_b^{\ast}\tan\b + \mu)$ in Eq.~(\ref{eq:G1}).

In the case $M_{\tilde{Q}} < M_{\tilde{D}}$, 
we have $(\ti b_1, \ti t_1)\sim (\ti b_L, \ti t_R)$ 
(since also $M_{\tilde{Q}} > M_{\tilde{U}}$) and 
hence $C^H_{\ti t_1\ti b_1}\sim m_t (A_t\cot\beta+\mu^{\ast})$  
(see Eq.~(\ref{eq:G})). Therefore $B(\ti b_1\to H^-\ti t_1)$
is nearly independent of $\varphi_ {A_b}$ which leads to
contour lines approximately
parallel to the $\varphi_{A_b}$-axis. 
In this case, however, nearly the whole
parameter space (i.e.\ the region with $|A_b| \lesssim 800$~GeV) shown in
Fig.~\ref{cpsbottom} (a) is excluded by the limit 
$B(b\to s \gamma) > 2.0 \times 10^{-4}$.

In Fig.~\ref{cpsbottom} (b) we show the contours of
$B(\tilde{b}_1 \to H^- \tilde{t}_1)$ as a function of $\varphi_{A_b}$
and $\varphi_{A_t}$ for $|A_t| =  |A_b| = 600$~GeV and the other
parameters (except $\varphi_{A_t}$) as in Fig.~\ref{cpsbottom} (a).
As can be seen, the $\varphi_{A_b}$-$\varphi_{A_t}$
correlation is even stronger than in Fig.~\ref{cpcharg} although it
has the same origin as that of 
$B(\tilde{t}_1 \to \tilde{\chi}^+_1 b)$.
Note that in the given scenario with $m_{H^\pm}=150$~GeV
the constraint on $B(b\to s \gamma)$ is only fulfilled
for a limited range of $\varphi_{A_t}$.
The case $M_{\tilde{Q}} < M_{\tilde{D}}$ is excluded because
$B(b\to s \gamma)$ is smaller than $2.0 \times 10^{-4}$ for this case.
Moreover we want to remark that even for small $\tan\beta$
the $\ti b_{1,2}$ decay branching ratios
can be somewhat sensitive to $\varphi_{A_{t,b}}$ 
and $\varphi_{\mu}$ \cite{short}.

\begin{figure}[t]
\centering
\begin{picture}(16,5.5)

\put(0,0.4){\epsfig{file=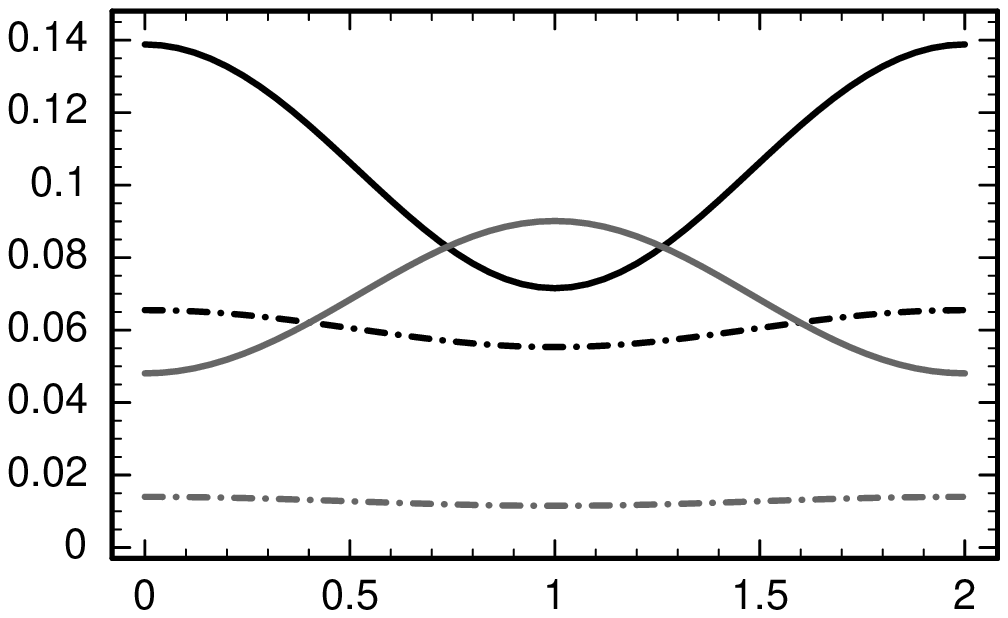,scale=.75}}    
\put(0.2,5.2){$B$}
\put(6.5,0.1){$\varphi_{A_b}/\pi$}
\put(3.95,5.3){(a)}

\put(8.4,0.4){\epsfig{file=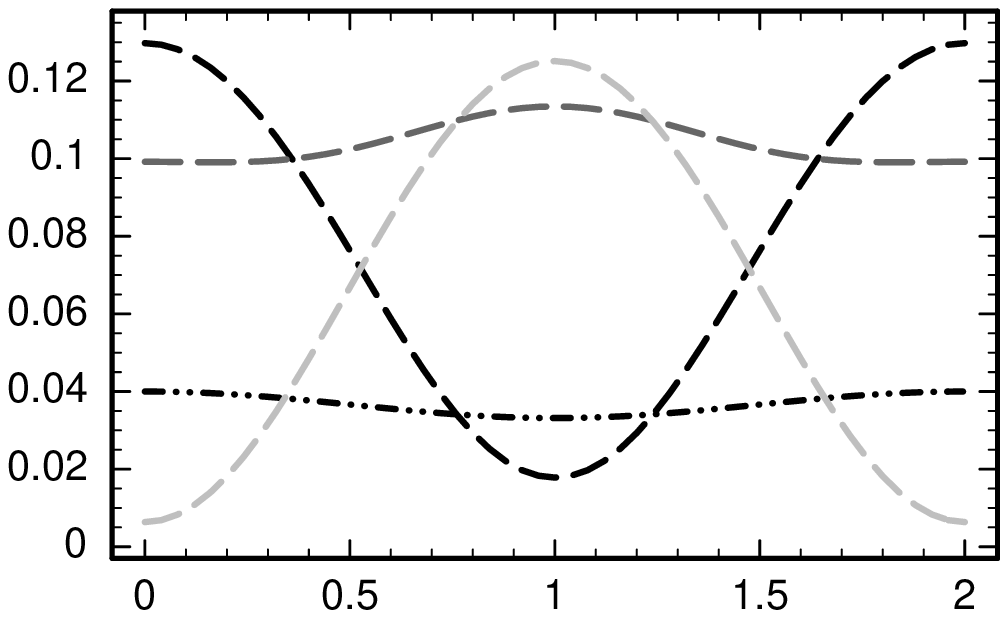,scale=.75}}    
\put(8.6,5.1){$B$}
\put(14.9,0.1){$\varphi_{A_b}/\pi$}
\put(12.35,5.3){(b)}

\end{picture}

\caption{\label{brphiAbSb2} $\varphi_{A_b}$ dependences of the
branching ratios of the bosonic decays
(a) $\tilde{b}_2 \to W^- \tilde{t}_{1/2}$ (dashdotted, black/gray),
$\tilde{b}_2 \to H^- \tilde{t}_{1/2}$ (solid, black/gray)
and (b) $\tilde{b}_2 \to Z \tilde{b}_1$ (dashdotdotted),
$\tilde{b}_2 \to H_{1/2/3} \tilde{b}_1$ (long dashed, black/gray/light
gray)
for $\tan\beta = 30$,
$M_2=200$~GeV, $|\mu| = 350$~GeV, $|A_b|=|A_t|=600$~GeV,
$\varphi_\mu=\varphi_{A_t}=\pi$, $\varphi_\mathrm{U(1)}=0$, 
$m_{\tilde{b}_1}=350$~GeV, $m_{\tilde{b}_2}=700$~GeV,
$m_{\tilde{t}_1}=170$~GeV and $m_{H^\pm}=150$~GeV, assuming
$M_{\tilde{Q}}<M_{\tilde{D}}$.}
\end{figure}

In case of the $\tilde{b}_2$ decays more decay channels are open. In
Fig.~\ref{brphiAbSb2} we show the branching ratios for the bosonic
decays $\tilde{b}_2 \to W^- \tilde{t}_{1,2},
Z \tilde{b}_1,  H^- \tilde{t}_{1,2}$ 
and $H_{1,2,3} \tilde{b}_1$ as a function of
$\varphi_{A_b}$ for $|\mu| = 350$~GeV, $|A_b|=|A_t|=600$~GeV,
$\varphi_\mu=\varphi_{A_t}=\pi$ and $\varphi_\mathrm{U(1)}=0$,
assuming $M_{\tilde{Q}}<M_{\tilde{D}}$.
The branching ratios of the fermionic decays are nearly independent of
$\varphi_{A_b}$ in this scenario.
The phase dependence of $\Gamma(\tilde{b}_2 \to W^- \tilde{t}_{1,2})$ and
$\Gamma(\tilde{b}_2 \to Z \tilde{b}_1)$ is caused solely by the phase
dependence of the squark mixing angles $\theta_{\tilde{b}}$ and
$\theta_{\tilde{t}}$ which is very weak in this scenario.
The strong $\varphi_{A_b}$ dependence of 
$\Gamma(\tilde{b}_2 \to H^- \tilde{t}_{1,2})$ is caused by the term
$m_b\,(A_b^{\ast}\tan\b + \mu)$ in the coupling $C^H_{\st\,\sb}$
(Eqs.~(\ref{eq:G}) and (\ref{eq:G1})).
As $\ti b_2\sim \ti b_R$ and 
$\varphi_{\ti t}\simeq \varphi_{A_t}=\pi$ in this case, 
the dominating term in the coupling $C^H_{\ti t\ti b}$ is
$(\mathcal{R}^{\tilde{t}} G)_{12} \simeq - e^{-i\varphi_{A_b}}
\cos\theta_{\tilde{t}} m_b |A_b| \tan\b + 
2 \sin\theta_{\tilde{t}} m_t m_b/\sin 2\b$
for $\tilde{b}_2 \to H^- \tilde{t}_1$ and
$(\mathcal{R}^{\tilde{t}} G)_{22} \simeq - e^{-i\varphi_{A_b}}
\sin\theta_{\tilde{t}} m_b |A_b| \tan\b - 
2 \cos\theta_{\tilde{t}}  m_t m_b/\sin 2\b$
for $\tilde{b}_2 \to H^- \tilde{t}_2$. 
Therefore, $B(\tilde{b}_2 \to H^- \tilde{t}_1)$ and 
$B(\tilde{b}_2 \to H^- \tilde{t}_2)$ behave like 
$(1+\cos\varphi_{A_b})$ and $(1-\cos\varphi_{A_b})$, respectively.
As in the example for the $\tilde{t}_2$ decays (Fig.~\ref{brphiAtSt2}) the
$\varphi_{A_b}$ dependence of 
$B(\tilde{b}_2 \to H_i \tilde{b}_1)$ $(i=1,2,3)$ is mainly due to the
phase factors explicitely appearing in Eq.~(\ref{eq:HLRcoupb})
whereas the $\varphi_{A_b}$ dependence of the $O_{ij}$ is less
pronounced.
Furthermore, there is only a small mixing in the bottom squark sector with
$\tilde{b}_2 \approx \tilde{b}_R$ and $\tilde{b}_1 \approx \tilde{b}_L$
in this scenario.
Hence the phase dependence of $B(\tilde{b}_2 \to H_i \tilde{b}_1)$ can
be explained by the phase dependence of 
$C(\sb_{L}^{\dagger} H_i \sb_{R})$ (Eq.~(\ref{eq:HLRcoupb})).
It turns out that
$H_1$ and $H_3$ are nearly CP-even Higgs bosons ($\phi_{1,2}$) with 
$O_{3i} \approx 0$ and $|\mu| O_{2i} \approx |A_b| O_{1i}$ $(i=1,3)$, 
which results in the pronounced $\varphi_{A_b}$ dependence of 
$B(\tilde{b}_2 \to H_{1,3} \tilde{b}_1)$.
$H_2$ is mainly a CP-odd Higgs boson $(a)$ with 
$O_{12} \approx O_{22} \approx 0$ and 
$\sin\beta |A_b| \gg \cos\beta |\mu|$, resulting in the weak
$\varphi_{A_b}$ dependence of $B(\tilde{b}_2 \to H_2 \tilde{b}_1)$.

We have analyzed the $\tilde{b}_2$ decay branching ratios also for
$M_{\tilde{Q}} > M_{\tilde{D}}$. 
The $\varphi_{A_b}$ dependence of 
$B(\tilde{b}_2 \to Z \tilde{b}_1)$ and 
$B(\tilde{b}_2 \to H_{1,2,3} \tilde{b}_1)$ are similar 
to those in Fig.~\ref{brphiAbSb2} (b), but they
are smaller by a factor of $\sim$ 3. 
The other branching ratios are nearly independent of
$\varphi_{A_b}$. 
However, for the scenario of Fig.~\ref{brphiAbSb2} the case of
$M_{\tilde{Q}}>M_{\tilde{D}}$ is excluded because
$B(b\to s \gamma)$ is smaller than $2.0 \times 10^{-4}$ for this case.

The $\varphi_\mathrm{U(1)}$ dependence of the partial decay widths 
and branching ratios in the scenario of Fig.~\ref{brphiAbSb2} with
$M_{\tilde{Q}} < M_{\tilde{D}}$ is very weak for the same reason as in
the scenario of Fig.~\ref{sbottomphiAb} with 
$M_{\tilde{Q}} > M_{\tilde{D}}$ for the decays of the $\tilde{b}_1$.

\section{Parameter Determination}
 
We now study to which extent one can extract
the underlying parameters from measured masses, branching ratios and
cross sections. 
Having in mind that the squark masses are relatively large in the scenarios
considered, we assume
the following situations:
(i) A high luminosity linear collider like TESLA can measure the masses of
charginos, neutralinos and the lightest neutral Higgs boson with
high accuracy \cite{tdr,Martyn:1999tc}. 
In the case that the squarks and the
heavier Higgs bosons have masses below
500 GeV, their masses can be measured with an error of
1\% and 1.5 GeV, respectively.
(ii) For SUSY particles with masses larger than 500 GeV 
their masses can be measured at a 2 TeV $e^+ e^-$ collider, 
such as CLIC. The masses of
heavy Higgs bosons and squarks can be measured with an error of 1\% and
3\%, respectively \cite{albert,CLIC}. For the production 
we can get an  $e^-$ beam polarization of $P_-=$ 0.8 and
 an $e^+$ beam polarization of $P_+=$ 0.4.
(iii) The gluino mass can be measured at the LHC with an error of 3\%
\cite{albert}.
(iv) $m_t$ can be measured with an error of 0.1~GeV.
In this case this error can be neglected in the fitting procedure
\cite{Heinemeyer:2003ud}. 
We assume that the error on $m_b$ can also be neglected.
(v) The branching ratio of $b\to s \gamma$ can be measured within an
error of $0.4 \times 10^{-4}$.

We do not take into account additional information from LHC
about the $\st_i$ and $\sb_i$ systems, because the amount of
information available strongly depends on the scenario realized in
nature \cite{Polesello}.
For example in the SPS1a scenario the decay channel
$\st_i \to b + \tilde{\chi}^+_1$ cannot be identified experimentally,
because the chargino decays into a scalar tau to practically 100\%.
$\st_i$ and $\sb_i$ production at LHC will probably not
give enough information about the stop and sbottom mixing angles.
Moreover, the formulae for the production cross sections at LHC,
which exist in the literature are for the real case only and do not
include complex phases, which might be important for the one-loop
corrections.
This is the main reason why we
did not consider LHC data for the stop and sbottom systems, but data
from CLIC. To clarify the situation at LHC concerning the scenarios
we considered would require additional theoretical work including
complex phases and further Monte Carlo studies, which are beyond the
scope of this paper.

Our strategy for the parameter determination is as follows:
\begin{enumerate}
\item Take a specific set of values of the underlying MSSM
parameters. 
\item Calculate the masses of
$\tilde t_i$, $\tilde b_i$, $\tilde \chi^0_j$, $\tilde \chi^\pm_k$,
$H_{\ell}$, the production cross sections for $e^+ e^- \to \tilde t_i
\bar{\tilde t}_j$, and $e^+ e^- \to \tilde b_i \bar{\tilde b}_j$, and 
the branching
ratios of the $\tilde t_i$ and $\tilde b_i$ decays.
\item Regard these calculated values as real experimental data with definite
errors.
\item Determine the underlying MSSM parameters and their errors from the
``experimental data'' by a fit using the program MINUIT 
\cite{James:1975dr}. 
\end{enumerate}

We consider two scenarios in the following, one with 
small $\tan\beta$ and
one with large $\tan\beta$. The small $\tan\beta$ scenario is characterized
by: $M_{\ti D} =$   169.6~GeV, $M_{\ti U} =$   408.8~GeV, 
$M_{\ti Q} =$   623.0~GeV,
$|A_t| = |A_b| = $ 800 GeV, 
$\varphi_{A_t} = \varphi_{A_b} = \pi/4$, $\varphi_{U(1)}=0$,
$M_2 =$  300~GeV, $\mu = -350$~GeV, $\tan \beta = 6$, 
 $m_{\tilde g} =$ 1000~GeV, and $m_{H^\pm} =$  900~GeV.
(Here we do not assume the unification relation between
$m_{\tilde g}$ and $M_2$.)
The resulting masses and their assumed experimental errors are:
$m_{\tilde \chi^\pm_1}= (  278.5\pm     0.2 )$~GeV,
$m_{\tilde \chi^\pm_2}= (  384.5\pm     0.3 )$~GeV,
$m_{\tilde \chi^0_1}= (  148.7\pm     0.3 )$~GeV,
$m_{\tilde \chi^0_2}= (  277.8\pm     0.5 )$~GeV,
$m_{\tilde \chi^0_3}= (  359.1\pm     0.3 )$~GeV,
$m_{\tilde \chi^0_4}= (  382.0\pm     0.7 )$~GeV,
$m_{H_1}= (  115.47\pm     0.05 )$~GeV,
$m_{H_2}= (  896.5\pm     9.0 )$~GeV,
$m_{H_3}= (  897.1\pm     9.0 )$~GeV,
$m_{\tilde t_1}= ( 350.0\pm     3.5 )$~GeV,
$m_{\tilde t_2}= ( 700.0\pm    21.0 )$~GeV,
$m_{\tilde b_1}= ( 170.0\pm     1.7 )$~GeV,
and $m_{\tilde b_2}= (  626.0\pm    19.0 )$~GeV.
Moreover, we find $B(b \to s \gamma)=     3.6\times 10^{-4}$.
The corresponding top squark and bottom squark branching ratios are given in
Tab.~\ref{tab:brfit}.
The large $\tan\beta$ scenario is specified by
 $M_{\ti D} =$   360.0~GeV,
 $M_{\ti U} =$   198.2~GeV,
 $M_{\ti Q} =$   691.9~GeV,
$|A_t| = $ 600 GeV, 
$\varphi_{A_t} = \pi/4$,
$|A_b| = $ 1000 GeV, 
$\varphi_{A_b} = 3 \pi/2$,
$\varphi_{U(1)} = 0$,
 $M_2 =$  200~GeV,
$\mu = -350$~GeV,
  $\tan \beta =$   30,
 $m_{\tilde g} =$ 1000~GeV, and
  $m_{H^\pm} =$  350~GeV.
 The resulting masses and their assumed errors are:
$m_{\tilde \chi^\pm_1}= (  188.2\pm     0.5 )$~GeV,
$m_{\tilde \chi^\pm_2}= (  374.2\pm     0.9 )$~GeV,
$m_{\tilde \chi^0_1}= (   98.2\pm     0.6 )$~GeV,
$m_{\tilde \chi^0_2}= (  188.2\pm     0.9 )$~GeV,
$m_{\tilde \chi^0_3}= (  358.5\pm     0.9 )$~GeV,
$m_{\tilde \chi^0_4}= (  371.6\pm     2.0 )$~GeV,
$m_{H_1}= (  113.63\pm     0.05 )$~GeV,
$m_{H_2}= (  340.7\pm     1.5 )$~GeV,
$m_{H_3}= (  341.1\pm     1.5 )$~GeV,
$m_{\tilde t_1}= (  210.0\pm     2.1 )$~GeV,
$m_{\tilde t_2}= (  729.0\pm    22.0 )$~GeV,
$m_{\tilde b_1}= (  350.0\pm     3.5 )$~GeV,
and $m_{\tilde b_2}= (  700.0\pm    21.0 )$~GeV.
Moreover, we have $B(b \to s \gamma)=     4.4\times 10^{-4}$.
The corresponding top squark and bottom squark branching ratios are given in
Tab.~\ref{tab:brfit}.
We have chosen a relatively small
$\sb_1$ mass in the small $\tan\beta$ scenario and a relatively small
$\st_1$ mass in the large $\tan\beta$ scenario.
As a result of this
in the two scenarios considered the observables in the $\st_i$
and $\sb_i$ sectors are sufficient to determine all $\st_i$ and
$\sb_i$ parameters.

\begin{table}[t]
\caption{Decay branching ratios (in $\%$) for top squarks and 
bottom squarks in the   
two considered scenarios.
  Corresponding values of the underlying MSSM parameters 
are given in the text.}
 \label{tab:brfit}
 \begin{center}
 \begin{tabular}{|c||c|c||c|c||c|c||c|c|}
 \hline
 & \multicolumn{4}{|c||}{scenario with $\tan\beta=6$}
 & \multicolumn{4}{|c|}{scenario with $\tan\beta=30$} \\
 channel & $\tilde t_1$   & $\tilde t_2$ 
         & $\tilde b_1$   & $\tilde b_2$ 
         & $\tilde t_1$   & $\tilde t_2$ 
         & $\tilde b_1$   & $\tilde b_2$ \\
 \hline
 $q \tilde \chi^0_1$ &    
 66.4 &
  1.6 &
100 &
  0.6 &
  0 & 
  0.6 &
 63.5 &
  0.6 \\
 $q \tilde \chi^0_2$ &    
  0 & 
  7.5 &
  0 & 
  8.7 &
  0 & 
  8.5 &
 36.1 &
 10.3 \\
 $q \tilde \chi^0_3$ &    
  0 & 
 13.1 &
  0 & 
  0.3 &
  0 & 
 11.1 &
  0 & 
  4.6 \\
 $q \tilde \chi^0_4$ &    
  0 & 
  6.6 &
  0 & 
  2.4 &
  0 & 
  8.7 &
  0 & 
  4.6 \\
 $q' \tilde \chi^\pm_1$ & 
 33.1 &
 19.2 &
  0 & 
  9.7 &
100 &
 22.5 &
  0 & 
 14.1 \\
 $q' \tilde \chi^\pm_2$ & 
  0 & 
  1.6 &
  0 & 
 21.0 &
  0 & 
  6.8 &
  0 & 
 24.2 \\
 $W^\pm \tilde q_1'$ &    
  0.5 &
  0.3 &
  0 & 
 56.8 &
  0 & 
  3.1 &
  0.4 &
 27.1 \\
 $H^\pm \tilde q_1'$ &    
  0 & 
  0 & 
  0 & 
  0 & 
  0 & 
  7.7 &
  0 & 
  6.4 \\
 $Z \tilde q_1$ &       
  $-$ & 
 26.9 &
  $-$ & 
  0.2 &
  $-$ & 
 13.1 &
  $-$ & 
  1.5 \\
 $H_1 \tilde q_1$ &     
  $-$ & 
 23.4 &
  $-$ & 
  0.2 &
  $-$ & 
 12.7 &
  $-$ & 
  1.4 \\
 $H_2 \tilde q_1$ &     
  $-$ & 
  0   &
  $-$ & 
  0   &
  $-$ & 
  2.8 &
  $-$ & 
  2.7 \\
 $H_3 \tilde q_1$ &     
  $-$ & 
  0   &
  $-$ & 
  0   &
  $-$ & 
  2.4 &
  $-$ & 
  2.7 \\
 \hline
 \end{tabular}
 \end{center}
\end{table}

We have taken the relative errors of 
chargino and neutralino masses from \cite{tdr,Martyn:1999tc}, which
we rescale according to our scenario;
in case of $\tan \beta = 30$ we have taken into account an
additional factor of 3 for the errors (relatively to $\tan\beta=6$)
due to the reduced efficiency in case of multi $\tau$ final states 
from decays of charginos and neutralinos as
indicated by the studies in \cite{Nojiri:1996fp}.

A detailed Monte Carlo study of the $\st_1$ production
$e^+e^- \to \st_1 \bar{\st_1}$ and the
$\st_1$ decays $\st_1 \to c\tilde{\chi}^0_1$ and 
$\st_1 \to b\tilde{\chi}^+_1$ at TESLA ($\sqrt{s}=500$~GeV and
$\mathcal{L}=500~\mathrm{fb}^{-1}$)
has been performed in \cite{Keranen:py}
for real MSSM parameters.
These results cannot directly be used for our error analysis,
because we consider additional $\st_1$ and $\st_2$ decays.
To the best of our knowledge no Monte Carlo
studies exist which include all of the  $\st_1$ and $\st_2$ decays
considered in our analysis.
Therefore, we have taken only statistical errors for the production
cross sections and branching ratios by calculating the corresponding
number of events for the decay $\st_1 \to X$ as
\begin{equation}
N = 2 \mathcal{L} 
 \left[ \sigma(\st_1\bar{\st_1}) + \sigma(\st_1\bar{\st_2}) \right]
 B(\st_1 \to X)
\end{equation} 
and analogously for $\st_2$, $\sb_1$ and $\sb_2$ decays.
For definiteness we take an integrated luminosity
$\mathcal{L} = 1~\mathrm{ab}^{-1}$ at a c.m.s.\ energy $\sqrt{s} =
2$~TeV (i.e.\ at CLIC).
We do not take systematic experimental
errors for the cross sections and branching ratios into account
since we are not aware of any study considering the systematic errors.
Instead we have doubled the statistical errors obtained above.
The evaluation of the systematic experimental errors would require
further Monte Carlo studies for a specific linear collider which, however, 
are beyond the scope of our paper.

For the determination of the squark parameters we have used the
information obtained from the measurement of the squark masses at 
threshold and the squark production cross sections  at $\sqrt{s}=2$~TeV for two
different $(e^-,e^+)$ beam polarizations $(P_-,P_+) = (0.8, -0.4)$ and
$(P_-,P_+) = (-0.8, 0.4)$. Here we have assumed that 
a total effective luminosity of 1 ab$^{-1}$ is available for
each choice of polarization. The cross section measurements are important
for the determination of $|\cos \theta_{\tilde t}|^2$ and
 $|\cos \theta_{\tilde b}|^2$ as can
be seen from Eq.~(\ref{eq:cij}) and the formulas for the cross
sections in \cite{sferm}.
In the numerical evaluation
of the squark production cross sections
we have included initial state radiation according to \cite{ISR}. 
In addition we have used the information from all 
branching ratios in Table~\ref{tab:brfit} 
with the corresponding statistical errors.
These branching ratios together with the masses and cross
sections form an over-constraining system of observables for the
underlying parameters 
$M_{\ti D}^2$, $M_{\ti U}^2$, $M_{\ti Q}^2$, 
${\rm Re}(A_t)$,  ${\rm Im}(A_t)$,  ${\rm Re}(A_b)$,  ${\rm Im}(A_b)$, 
${\rm Re}(M_1)$, ${\rm Im}(M_1)$, $M_2$, ${\rm Re}(\mu)$, 
${\rm Im}(\mu)$, $\tan \beta$, 
$m_{\tilde g}$, and $m_{H^\pm}$. The latter two 
enter the formulas for the neutral Higgs masses and mixing. 
We determine these parameters and their errors from the ``experimental
data'' on these observables by a
least-square fit.
The results obtained are shown in Table~\ref{tab:parfit}. 
Note that the sign ambiguity for the
imaginary parts of the parameters 
is due to the fact that we consider CP-even observables. 
This ambiguity can in principle be resolved by considering appropriate CP-odd
observables (as proposed in \cite{Aoki:1998cq} -- \cite{Bartl:2003ck})
in the analysis.
As one 
can see, all parameters except $A_b$ can be determined rather precisely.
$\tan\beta$ can be determined with
an error of about 3\% in both scenarios.  
The relative error of the squark mass parameters squared is in the
range of 1\% to 2\%. $A_t$ can be measured within an error of 
2 -- 3\% independently of
$\tan\beta$. The reasons for this are: (i) the mixing angle in
the top squark sector, which can be measured rather precisely 
using polarized $e^{\pm}$ beams,
depends strongly on $A_t$ and (ii) $A_t$ influences strongly the corrections
to the mass of the lightest Higgs boson. 
The situation for $A_b$ is considerably worse: in case of small $\tan\beta$
one gets only an order of magnitude estimate. The reason is that
both the bottom squark mixing angle and the bottom squark couplings depend
only weakly on  $A_b$ for small $\tan\beta$. In case of large 
$\tan\beta$ the situation improves somewhat in particular for
the imaginary part of $A_b$. The main sources of information on $A_b$ are
the branching ratios of 
the decays of the heavier bottom squark into a Higgs boson plus the
lighter bottom
squark because the corresponding couplings depend significantly on $A_b$
(see Eqs.~(\ref{eq:HLLcoupb}) -- (\ref{eq:HLRcoupb})). From this we conclude
that the situation for $A_b$ improves in scenarios
where these branching ratios are large. An additional source of information
could be the polarization information of the fermions in bottom squark decays
as proposed in \cite{Boos:2003vf}.
We have found that the analogous fit procedure for
real MSSM parameters gives a larger value for $\chi^2$:
$\Delta\chi^2=286.6$ for the scenario with $\tan\beta=6$ and 
$\Delta\chi^2=22.5$ for the scenario with $\tan\beta=30$.
In Table~2 most of the central values of the fitted
parameters are the same as their input values because we have
taken the observables calculated from the input parameters
as ``experimental data''. We have
checked that a shift within $1\sigma$ of the ``experimental data''
leads to almost no change of the errors of the parameters.

\begin{table}[t]
\caption{Extracted parameters from the ``experimental data'' of the
 masses, production cross sections and decay branching ratios of
 $\tilde t_i$ and $\tilde b_i$. The original parameters for each scenario
are given in the text.}
 \label{tab:parfit}
\begin{center}
 \begin{tabular}{|c|c|c|}
 \hline
scenario      &    $\tan\beta=6$ scenario & $\tan\beta=30$ scenario \\ \hline
$M^2_{\tilde{D}}$ &    $(2.88 \pm 0.06) \times 10^4$ & $(1.30 \pm  0.02) \times 10^5$ \\
$M^2_{\tilde{U}}$ &    $(1.67 \pm 0.04) \times 10^5$ & $(3.93 \pm  0.12) \times 10^4$ \\
$M^2_{\tilde{Q}}$ &    $(3.88 \pm 0.04) \times 10^5$ & $(4.79 \pm  0.04) \times 10^5$ \\
Re($A_t$)        &    $565.0 \pm 13.0     $    &  $424.0 \pm 14.0$   \\
Im($A_t$)        &    $\pm 566.0 \pm 14.0 $    &  $\pm 425.0  \pm  15.0 $  \\
Re($A_b$)        &    $620.0 \pm   190.0$     &  $6.5 \pm  420.0$  \\
Im($A_b$)        &    $\pm 230.0 \pm   580.0$ &  $\pm 999.0 \pm 52.0$ \\ \hline
Re($M_1$)        &    $149.3 \pm 0.3$  &    $99.6 \pm   0.6$ \\
Im($M_1$)        &    $1.0 \pm   1.5$ &    $-0.5 \pm  2.8$  \\
$M_2$            &    $300.0 \pm 0.4$ &     $200.0 \pm   0.5$  \\
Re($\mu$)        &   $-350.0 \pm 0.3$ &    $-350.0 \pm  0.6$ \\
Im($\mu$)        &   $-0.02 \pm  0.9$  &    $1.5  \pm  5.0$  \\
$\tan\beta$      &    $6.0 \pm  0.2$   &    $30.0 \pm 0.8$ \\ \hline
$m_{\tilde g}$   &    $1000.0 \pm 30$  &    $1000.0 \pm 30$ \\
$m_{H^{\pm}}$    &    $900.0 \pm  5.0$ &    $350.0 \pm   0.8$ \\
\hline
 \end{tabular}
\end{center}
\end{table}

The results presented in Table~\ref{tab:parfit} 
depend clearly
on the assumed experimental errors which have been summarized in the beginning
of this section. 
It is clear that further detailed Monte Carlo studies including
experimental cuts and detector simulation are necessary to determine more
accurately the expected experimental errors of the observables
for our scenarios, in
particular the errors of the top squark and bottom squark decay
branching ratios.
Such a study is, however,
beyond the scope of this paper.
Furthermore, an additional source of uncertainty is the
theoretical error due to higher order corrections etc.\
\cite{loopdecay, Guasch:2002ez}. We have not taken into account these effects
because most formulae given in the literature are only for real parameters.
Instead we have studied how our results
for the errors of the fundamental parameters are changed when the
experimental errors of the various observables are changed: we have redone
the procedure doubling the errors of the masses and/or
branching ratios and/or cross sections. We find that the errors
of all parameters are approximately doubled if all experimental
errors are doubled.
Moreover, in this way we can see to which observables an individual parameter
is most sensitive.
We find that precision on the top squark parameters 
$A_t, M^2_{\ti Q}$ and $M^2_{\ti U}$ is sensitive to 
the accuracy of the top squark mass measurement
at the threshold as well as to the precision of 
the measurement of the total top squark pair production cross sections
in the continuum using polarized $e^{\pm}$ beams. 
The error of $A_t$ is also very sensitive to
the error of the lightest Higgs boson mass due to the large top squark loop
corrections. The precision on the parameters $M^2_{\tilde D}$ 
and $M^2_{\tilde Q}$ is sensitive to 
the accuracy of the bottom squark mass measurement.
The accuracy of $A_b$ is most sensitive to the precision of the
measurements of the branching ratios for the bottom squark (and top squark)
decays into Higgs bosons. The precision of $\mu$ is more sensitive to the 
errors
of chargino and neutralino masses than to the errors of the top squark and
bottom squark observables.
In the case of large $\tan \beta$, the precision of $\tan \beta$ depends 
to some extent
on the precision of the bottom squark pair production
cross sections and to a lesser extent
also on that of the bottom squark decay branching ratios. 

For the determination of the $\st_i$ and $\sb_i$ parameters
the measurements of the branching ratios of the squark decays into
Higgs bosons together with those of the squark mixing angles from the
production cross sections are important. Therefore, we
need to obtain information about $\st_1$, $\st_2$, $\sb_1$ and
$\sb_2$ production and decays separately.
This can be achieved at a Linear Collider by suitable choices of the
c.m.s.\ energy.
We remark that,
in the case $m_{\ti t_2}, m_{\ti b_2}\gsim 500$~GeV the measurements of
the cross sections, masses and branching ratios of $\ti t_2$ and $\ti b_2$ 
at an $e^+ e^-$ linear collider with $\sqrt{s}=2$~TeV are necessary 
for the determination of $A_t$ and $A_b$, otherwise this 
might not be possible.
However, additional information from the LHC on the $\st_i$
and $\sb_i$ masses and some of the decay channels would certainly
improve the situation.
In the error estimate presented here we have assumed that many
decay channels of the $\st_i$ and $\sb_i$ are open. If this is not the
case, then the missing information could be obtained by studying the
decay properties of the heavier charginos, neutralinos and Higgs
bosons into $\st_i$ and $\sb_i$.

\section{Summary}

In this paper we have studied the decays of top squarks $\tilde{t}_i$ and
bottom squarks $\tilde{b}_i$ in the MSSM with
complex parameters $A_t$, $A_b$, $\mu$ and $M_1$. We have taken into
account the explicit CP violation in the Higgs sector induced by
$\tilde{t}_i$ and $\tilde{b}_i$ loops in the case
$A_{t,b}$ and $\mu$ are complex. We have presented numerical
results for the fermionic and bosonic decay branching ratios
of $\tilde{t}_i$ and
$\tilde{b}_i$ ($i = 1,2$). We have analyzed their MSSM parameter
dependence, in particular the dependence on the CP phases
$\varphi_{A_t}$, $\varphi_{A_b}$, $\varphi_\mu$ and
$\varphi_\mathrm{U(1)}$.
We have found that the experimental data of the branching
ratio of the
decay $b\to s\gamma$ can lead to considerable restrictions
on the MSSM parameter space.
In the case of $\tilde{t}_i$ decays the strong dependence on $\varphi_{A_t}$
and $\varphi_\mu$ is due to the phase dependence of the mixing
angle $\theta_{\tilde{t}}$, of the mixing phase factor
$e^{i\varphi_{\tilde{t}}}$ and of the Higgs couplings $G_{12}$
$(=C^H_{\ti b_R\ti t_L})$,
$G_{21}$ $(=C^H_{\ti b_L\ti t_R})$ and $C(\st_{L}^{\dagger} H_i \st_{R})$.
In the case of $\tilde{b}_i$ decays there can be strong
$\varphi_{A_b}$ dependence if $\tan\beta$ is large and the decays
into Higgs bosons are allowed.
If the parameters $A_t$, $A_b$, $\mu$ and $M_1$ are complex and
there is mixing between the CP-even and CP-odd Higgs bosons, the decay
pattern of $\tilde{t}_i$ and $\tilde{b}_i$ is even more complicated
than that in the case of real parameters. This could have important
implications for $\tilde{t}_i$ and $\tilde{b}_i$ searches at future
colliders and the determination of the underlying MSSM parameters.

We have also estimated what accuracy can be expected in the
determination of the underlying MSSM parameters by
a global fit of the observables (masses, branching ratios and production
cross sections) measured at typical linear colliders with polarized beams. 
We have
considered two scenarios with $\tan\beta = 6$ and $\tan\beta = 30$.
Under favorable conditions the fundamental MSSM parameters except $A_{t,b}$
can be determined with errors of 1\,\% to 2\,\%, assuming an
integrated luminosity of 1~ab$^{-1}$. The parameter $A_t$ can be
determined within an error of 2 -- 3\,\% 
whereas the error of $A_b$ is likely to be of the
order of 50\,\%.

\section*{Acknowledgements}

We thank M.~Battaglia, A.~De Roeck, H.~Eberl, M.~Kr\"amer,
W.~Majerotto, G.~Moortgat-Pick and G.~Weiglein for useful discussions.
K. H. appreciates valuable discussions with E. Berger, H. Haber,
G. Kane and P. Nath.
This work is supported by the `Fonds zur F\"orderung der
wissenschaftlichen For\-schung' of Austria, FWF Projects No.~P13139-PHY
and No.~P16592-N02
and by the European Community's Human Potential Programme
under contract HPRN-CT-2000-00149.
W.P.~has been supported by the Erwin Schr\"odinger fellowship 
No.~J2272 of the `Fonds zur F\"orderung der wissenschaftlichen 
Forschung' of Austria and partly by the Swiss 'Nationalfonds'.

\section*{Appendix}

\begin{appendix}

\section{Masses and mixing in the neutral Higgs sector}

In the complex MSSM the explicit CP violation in the Higgs sector is
mainly induced by $\tilde{t}$ and $\tilde{b}$ loops
resulting in a $3 \times 3$ neutral Higgs mass matrix with
a mixing of the CP-even Higgs bosons $\phi_1$
and $\phi_2$ and the CP-odd Higgs boson $a$.
At one-loop level the amount of mixing of CP-even and CP-odd
Higgs states is approximately proportional to
$\sin(\varphi_{A_{t,b}} + \varphi_\mu)$.
The three neutral mass
eigenstates are denoted as $H_i$ ($i=1,2,3$) with masses 
$m_{H_1} < m_{H_2} < m_{H_3}$ 
(following the notation of \cite{ref3}).
The real orthogonal mixing matrix in the neutral Higgs sector is
denoted by a $3 \times 3$ matrix $O$:
\begin{equation} \label{eq:nhiggsmix}
 \left(\begin{array}{c} H_1 \\ H_2 \\ H_3 \end{array}\right) =
  O^T \left(\begin{array}{c} \phi_1 \\ \phi_2 \\ a \end{array}\right),
\end{equation}
where $\phi_1$, $\phi_2$ and $a$ are related to the neutral entries
of the two Higgs doublet fields by 
$H^0_1 = 1/\sqrt{2} (v_1 + \phi_1 - i a_1)$,
$H^0_2 = 1/\sqrt{2} (v_2 + \phi_2 + i a_2)$
and $a = -\sin\beta a_1 + \cos\beta a_2$.
We take the parameter $\xi=0$ as in \cite{ref3}.
We have included the full one-loop corrections to the Higgs
mass eigenvalues $m_{H_i}$ and the mixing matrix $O_{ij}$ as
implemented in the program FeynHiggs2.0.2 \cite{feynhiggs}. We use these
results for $m_{H_i}$ and $O_{ij}$ in our tree-level formula for the
$\st_2$, $\sb_2$ decay widths (Eq.~(\ref{eq:gamneuthiggs})) and in the
constraint (ii).

\section{Chargino Masses and Mixing}
\label{app:char}

At tree-level the chargino mass matrix in the weak basis is given by 
\cite{susy,guha}
\begin{equation}
\mathcal{M}_C=\left(\begin{array}{ccc}
M_2 & \sqrt{2}m_W s_{\beta}\\[4mm]
\sqrt{2}m_W c_{\beta} & |\mu| e^{i\varphi_{\mu}}
\end{array}\right).
\end{equation}
$c_{\beta}$ and $s_{\beta}$ are
$\cos \beta$ and $\sin \beta$, respectively.
This complex $2 \times 2$ matrix is diagonalized by the unitary 
$2 \times 2$ matrices $U$ and $V$:
\begin{equation}\label{dia}
U^{\ast} \mathcal{M}_C V^{\dag}=
\mathrm{diag}(m_{\tilde\chi_1^{\pm}},m_{\tilde\chi_2^{\pm}})
, \hspace{2cm} 0\le m_{\tilde\chi_1^{\pm}} \le m_{\tilde\chi_2^{\pm}}.
\label{eq:massCH}
\end{equation}
We have neglected one-loop corrections to the chargino mass matrix
$\mathcal{M}_C$, as have been given in 
\cite{Guasch:2002ez, loopcharneutmass} for real parameters.

\section{Neutralino Masses and Mixing}
\label{app:neut}

At tree-level the neutralino mass matrix in the weak basis
$(\tilde B, \tilde W^3, \tilde H_1^0, \tilde H_2^0)$ is given as 
\cite{susy,guha}:
\begin{equation}
\mathcal{M}_N = \ \left( \begin{array}{cccc}
|M_1| e^{i\varphi_\mathrm{U(1)}} & 0 & -m_Z s_W c_{\beta} & m_Z
s_W s_{\beta}\\[3mm]
0 & M_2 & m_Z c_W c_{\beta} & -m_Z c_W s_{\beta}\\[3mm]
-m_Z s_W c_{\beta} & m_Z c_W c_{\beta} & 0 &
-|\mu| e^{i\varphi_{\mu}}\\[3mm]
m_Z s_W s_{\beta} & -m_Z c_W s_{\beta} &
-|\mu| e^{i\varphi_{\mu}} & 0
\end{array}\right),
\label{eq:massN}
\end{equation}
\\
where $\varphi_\mathrm{U(1)}$ is the phase of $M_1$, and
$c_W$ and $s_W$ are $\cos \theta_W$ and $\sin \theta_W$, respectively. 
This symmetric complex mass matrix is diagonalized by the
unitary $4 \times 4$ matrix $N$:
\begin{equation}
\label{eq:mixN}
N^{\ast}\mathcal{M}_N N^{\dag}\ = \mathrm{diag}(m_{\tilde\chi_1^0},\dots,
m_{\tilde\chi_4^0}),
\hspace{2cm} 0\le m_{\tilde\chi_1^0} \le \dots \le m_{\tilde\chi_4^0} \,.
\end{equation}
We have not included one-loop corrections to the neutralino mass matrix
$\mathcal{M}_N$, like those given in 
\cite{Guasch:2002ez, loopcharneutmass} for real parameters.

\end{appendix}

\end{document}